\documentclass[sn-mathphys,Numbered]{sn-jnl}

\usepackage{float}
\usepackage{placeins}
\usepackage[utf8]{inputenc}
\usepackage{amsmath}
\usepackage{chngcntr}
\usepackage{graphicx}%
\usepackage{multirow}%
\usepackage{amsmath,amssymb,amsfonts}%
\usepackage{amsthm}%
\usepackage{mathrsfs}%
\usepackage[title]{appendix}%
\usepackage{xcolor}%
\usepackage{textcomp}%
\usepackage{manyfoot}%
\usepackage{booktabs}%
\usepackage{algorithm}%
\usepackage{algorithmicx}%
\usepackage{algpseudocode}%
\counterwithin{equation}{section}

\usepackage{tabularx}   
\usepackage{listings}%
\usepackage{caption}      
\usepackage{matlab}
\usepackage{ragged2e}
\usepackage{setspace}
\theoremstyle{thmstyleone}%
\theoremstyle{thmstyletwo}%

\theoremstyle{thmstylethree}%

\begin{document}

\title[Article Title]{Constrained Dynamics on Eccentric Conic Orbits: Dirac–Bergmann and Hamilton–Jacobi Approaches}

\author[1]{\fnm{Alejandro Gabriel} \sur{Andarcia Caballero}}\email{alejandroandarcia@hotmail.com}

\author*[1]{\fnm{Jaime} \sur{Manuel Cabrera}}\email{jaime.manuel@ujat.mx}

\author*[1]{\fnm{Luis Gerardo} \sur{Romero Hernández}}\email{jorge.paulin@ujat.mx}

\author[1]{\fnm{Jorge Mauricio} \sur{Paulin-Fuentes}}\email{192a12006@egresados.ujat.mx}

\equalcont{These authors contributed equally to this work.}

\affil*[1]{\orgdiv{Divisi\'on Acad\'emica de Ciencias B\'asicas}, \orgname{Universidad Ju\'arez Aut\'onoma de Tabasco}, \orgaddress{\street{Km 1 Carretera
Cunduac\'an-Jalpa}, \city{ Cunduac\'an}, \postcode{86690}, \state{Tabasco}, \country{ M\'exico}}}




\abstract{

In this work, we investigate a Lagrangian model describing a particle constrained to move along non-degenerate conic sections, parameterized by the orbital eccentricity \( e \). In the non-relativistic regime, we apply the Dirac--Bergmann algorithm to identify a set of four second-class constraints, compute the corresponding Dirac brackets, and isolate the true physical degrees of freedom. This procedure yields a unified Hamiltonian treatment of circular (\( e = 0 \)), elliptical (\( 0 < e < 1 \)), parabolic (\( e = 1 \)), and hyperbolic (\( e > 1 \)) trajectories. We then extend the analysis to the relativistic case, where we observe a similar constraint structure and construct the associated Dirac brackets accordingly. Finally, using the Hamilton--Jacobi formalism, we identify a set of non-involutive constraints; by introducing generalized brackets, we restore integrability and derive the correct equations of motion. A comparative analysis of both formalisms highlights their complementary features and deepens our understanding of the dynamics governing particles restricted to conic geometries.

 }




\maketitle
\newpage
\section{Introduction}\label{sec1}

In classical mechanics, various formalisms allow us to describe the behavior of physical systems. Among the most fundamental are the Lagrangian and Hamiltonian approaches, which, together with Poisson brackets \cite{Lan,gold}, form the backbone of classical mechanics. These methods provide powerful tools for analyzing dynamical systems, revealing essential features such as degrees of freedom, equations of motion, symmetries, and conservation laws \cite{Deriglazov2017,Lemos_2018,mann2018lagrangian}.
However, their applicability is generally limited to regular systems—those whose constraints arise geometrically from the configuration of the system and are typically expressed as $\phi(q) = 0$\footnote{Example: The constraint of the ideal simple pendulum $\phi(q) = \phi(x, y) = x^{2} + y^{2} - l^{2} = 0$.}. In such cases, the equations of motion can be derived directly. But in more complex systems, inherent  constraints may emerge due to the structure of the phase space itself—such systems are associated with singular Lagrangians \cite{Teitelboim}. In these cases, the traditional Lagrangian and Hamiltonian frameworks are no longer sufficient on their own to fully determine the system’s dynamics, and additional tools are needed to deal with the constraints. 

The distinction between regular and singular systems is determined by the properties of the Hessian matrix \cite{hanson1964, sundermeyer1982constrained}, formed from the second derivatives of the Lagrangian with respect to the generalized velocities. If the determinant of this matrix is nonzero, the system is regular, allowing a unique inversion from velocities to momenta. Conversely, a vanishing determinant characterizes a singular system, where such inversion is not possible and constraints naturally arise as relations among the dynamical variables, particularly between coordinates and canonical momenta. To analyze such systems, one of the most widely used approaches is the Dirac–Bergmann formalism, developed by Paul Dirac \cite{Dirac} and Peter Bergmann \cite{Ber, Ber1}. This method systematically classifies constraints as primary, secondary, or tertiary, and further distinguishes them as first-class or second-class. This classification is key to constructing Dirac brackets, which play a central role in the system’s classical dynamics and lay the groundwork for canonical quantization \cite{Dirac}. 

A particularly illustrative application of these ideas is the analysis of a particle constrained to move along a conic section. Such systems exemplify the complexities inherent in singular dynamics, as the constraints encode geometric information that directly affects the structure of the phase space. Dirac’s canonical formalism has also played a central role in this context, providing powerful tools to study the dynamics of particles constrained to conic manifolds—such as relativistic trajectories \cite{oliveira2024relativistic} or motion on hyperboloidal surfaces \cite{najafizade2022dirac}. Due to the interplay between geometry and constraint structure, conic-constrained systems have attracted considerable attention in both classical and quantum contexts. To deal with the singularities these systems present, several formalisms have been developed. Among them, the BRST and BFV approaches have proven especially effective in the consistent quantization of systems with mixed first- and second-class constraints \cite{barbosa2018brst,thibes2021brst}, including generalized rotors and particles confined to curved geometries \cite{gomez2024symplectic}. Complementary to these quantization schemes, several strategies have been explored within Dirac’s formalism to handle the geometric complexity of conic constraints. A strategy within Dirac’s framework involves the introduction of dimensionless parameters (labeled A–F) that encode geometric features like orbital eccentricity. This allows all conic sections—circle, ellipse, parabola, and hyperbola—to be described within a unified Hamiltonian formalism \cite{barbosa2018gauge,caires2022general}. Other works adopt a more focused approach, examining the dynamics of particles confined to specific conic geometries, such as the circle \cite{fung2014dirac, scardicchio2002classical} or the ellipse \cite{nawafleh2013constrained}. 


Here, we adopt a unified approach: we express the conic constraint explicitly in terms of the eccentricity parameter $e$, which determines the geometry of the conic section \cite{landi2018linear}. We enforce the constraint using a Lagrange multiplier, introduced as an extra dynamical variable, thereby enlarging the configuration space. This strategy is widely used in the quantization of constrained systems \cite{najafizade2022dirac, caires2022general, nawafleh2013constrained, barbosa2018gauge, fung2014dirac, scardicchio2002classical, klauder1998coordinate, thibes2021brst}, and enables a consistent and unified treatment of all conic trajectories—circular, elliptical, parabolic, and hyperbolic—within a single formal framework.

While conceptually elegant, the Dirac–Bergmann algorithm can be technically intricate, requiring several logical steps and careful constraint management \cite{brown2022singular}. This complexity has motivated the development of alternative methods, such as the Hamilton–Jacobi (HJ) formalism \cite{guler1992canonical}. The extension of the HJ framework to systems with constraints aims to treat both the Hamilton–Jacobi equation and the constraints as a unified system of partial differential equations (PDEs) \cite{Nawafleh1, Nawafleh2, Na}. This approach differs from the conventional single HJ equation used in regular systems \cite{mann2018lagrangian}, and offers a promising route to address the integrability issues arising in singular systems.

The Hamilton–Jacobi formalism offers additional advantages across a wide range of theoretical contexts, including field theory, gauge theories, and general relativity. While significant progress has been made for systems with first-class constraints \cite{Gra, Gra1, Esca}, various approaches have also been proposed to handle second-class constraints \cite{BFT1, Batalin, Loran, Baleanu, rothe2010}. Importantly, the HJ formalism is also effective in finite-dimensional systems, which are simpler and often serve as test beds for ideas developed in field theories. In fact, it has been successfully applied to a range of mechanical models \cite{Bertin, Bertin2, romero2025singular}, demonstrating its flexibility and robustness in the study of constrained dynamics.

This work is structured as follows. In Section 2, we apply the standard Dirac–Bergmann Hamiltonian formalism to a particle constrained to move along a conic. The analysis, carried out in the non-relativistic regime, shows that the system is genuinely second-class, governed by a hierarchy of four second-class constraints. Within this framework, we also derive the equations of motion using the Dirac brackets, providing a consistent and systematic description of the constrained dynamics. Section 3 extends the discussion to the relativistic domain. Despite the additional algebraic complexity introduced by relativistic kinematics, the constraint structure remains unchanged, and the system retains its second-class nature. The Dirac bracket formalism continues to offer a coherent description of the reduced phase space, confirming its robustness under relativistic generalization. Sections 4 and 5 offer a complementary perspective through the Hamilton–Jacobi (HJ) formalism. In Section 4, we revisit the non-relativistic case and derive the associated Hamilton–Jacobi partial differential equations from the canonical Hamiltonian. Initially, the presence of non-involutive constraints prevents integrability. However, by introducing generalized brackets, we restore integrability and recover the equations of motion. These match those obtained from the Dirac–Bergmann approach, highlighting the consistency between the two methods. Section 5 follows a similar strategy for the relativistic extension. The formulation of the relativistic Hamilton–Jacobi equations involves more intricate constraint structures, but integrability is again achieved through generalized brackets. The resulting dynamics remain in full agreement with those derived from the Dirac formalism, underscoring the reliability of both approaches for analyzing second-class systems under relativistic conditions.

Appendices \ref{AppendixA} and \ref{AppendixB} present the detailed technical calculations that support the results discussed in Sections 2 and 3, respectively. Appendix \ref{AppendixC} provides a comprehensive symbolic implementation of the Dirac–Bergmann algorithm in MATLAB, which was used to reproduce, verify, and compare the analytical findings covered throughout Sections 2 to 5. This computational approach offers a clear and systematic framework to explore both the structure and effectiveness of the method.

\newpage

\section{Particle constrained on a non-degenerate conic section}


In this section, we apply the Dirac-Bergmann algorithm \cite{Teitelboim,hanson1964,sundermeyer1982constrained} to analyze the motion of a particle constrained to move on conic sections. We will systematically detail all steps and checks of the formalism, starting with the following Lagrangian \cite{landi2018linear}
\begin{eqnarray}
L= \frac{m}{2}(\dot{x}^{2}+\dot{y}^{2}) + \lambda\, T(x,y),
\label{2}
\end{eqnarray}
where \footnote{Where $\Omega = 1-e^2$ and  $\omega = a_x - e^2 u$.}
\begin{equation} 
    T(x,y) =
    x\,(\Omega\,x - 2\omega) + a_x^2 - e^2 u^2 + (y-a_y)^2.
    \label{3}
\end{equation} 
The quantity $\lambda$ is treated as an additional dynamical variable, serving solely to implement the constraint $T(x,y)$ as an equation of motion. The model described by equation (\ref{2}) inherently represents a second-class singular system from the Dirac-Bergmann (DB) perspective. Its constraint structure can be directly revealed by applying the Dirac constraint conservation algorithm, as demonstrated in the remainder of this section. Before proceeding with Dirac's formalism, we must confirm that we are indeed analyzing a singular system by evaluating its Hessian matrix
%
%
\begin{equation}
W_{ij}=\left(\frac{\partial^2 L}{\partial \dot{q}_n \partial \dot{q}_m}\right) , \quad q_{n}=(x,y,\lambda),
\label{3a}
\end{equation}
resulting in the matrix 
\begin{small}
\begin{eqnarray}
W=\left(
\begin{array}{cccc}
m & 0 & 0 & \\
0 & m & 0 \\
0 & 0 & 0 \\
\end{array}
\label{Hessianmatrix1}
\right),
\label{4}
\end{eqnarray}
\end{small}
which clearly has a determinant of zero, indicative of a singular system.

We proceed to calculate the canonical momenta 

\begin{equation}
P_n = \frac{\partial L}{\partial \dot{q}_n}. 
\label{4a}
\end{equation}

The conjugate momenta for this system are defined by 
\begin{eqnarray}
P_{x}  = m\dot{x}, \quad   P_{y}  = m\dot{y}, \quad P_{\lambda}  = 0,
\label{4b}
\end{eqnarray}
the absence of  $\dot{\lambda}$ in Lagrangian (1) directly leads to this natural primary constraint
\begin{equation}
\phi_1 = P_{\lambda} \approx 0.
\label{4c}
\end{equation}
To begin the Hamiltonian formulation, we compute the canonical Hamiltonian $H_{C}$ by the usual prescription
\begin{eqnarray}
    H_{C}=\dot{q}_{i}P^{i}-L(q_{i},\dot{q}_{i}), 
    \label{4d}   
\end{eqnarray}
and then express it solely in terms of the phase‑space variables $(q,p)$ \cite{Teitelboim, hanson1964,sundermeyer1982constrained,brown2022singular}. 

Although the degeneracy of the Lagrangian \eqref{2} prevents us from solving for all velocities $\dot{q}_{i}$ in terms of $(q,p)$, one show that the combination (\ref{4d}) depends only $(q,p)$. For our specific model, this procedure yields
%
%
\begin{equation}
    H_c = \frac{1}{2m}[{P_{x}}^2 + {P_{y}}^2]-\lambda\,T(x,y).
    \label{4e}
\end{equation}
However, since equation (\ref{2}) is singular, the Legendre transform fails to be invertible. This singularity gives rise to the primary constraint (\ref{4c}), which confines the Hamiltonian to the corresponding constraint surface. To systematically uncover all internal constraints of the system, we follow the Dirac–Bergmann algorithm \cite{Dirac,Teitelboim,hanson1964,sundermeyer1982constrained}.
As part of this procedure, we introduce a Lagrange multiplier 
$u_{1}$  and define the primary Hamiltonian $H_{p}$. This Hamiltonian is constructed by augmenting the canonical Hamiltonian
$H_{C}$ with the primary constraint multiplied by the Lagrange multiplier
%
%
%
\begin{eqnarray}
    H_{p}=H_{c}+u_{1}\phi_{1}.
    \label{5}
\end{eqnarray}
%
%
This extension ensures that the constraints are preserved under time evolution and facilitates the identification of possible secondary constraints. To guarantee that the primary constraints remain consistent throughout the system's dynamics, we invoke Dirac's consistency conditions~\cite{Dirac}. The procedure begins by evaluating the time evolution of each primary constraint using the primary Hamiltonian.

Specifically, we require the stationarity of the primary constraint $\phi_1$ under time evolution:
\begin{eqnarray}
     \dot{\phi_{1}}=\{\phi_{1},H_{p}\} \approx 0, 
     \label{5a}
\end{eqnarray}
This condition implies that the Poisson bracket of $\phi_1$ with the primary Hamiltonian $H_P$ must vanish weakly. If the resulting expression depends on a Lagrange multiplier, its value can be adjusted to satisfy the condition. Otherwise, the expression must vanish independently of the multipliers, leading to the emergence of a secondary constraint. This process is iterated for all newly generated constraints until closure is achieved—i.e., until no further constraints appear and all consistency conditions are satisfied.

For our system, the primary constraint is given by (\ref{4c}), and its time derivative reads
\begin{eqnarray}
    \dot{\phi_{1}}=\{\phi_{1},H_{p}\}=-T(x,y)\approx 0. 
    \label{6}
\end{eqnarray}
Consequently, we identify the corresponding secondary constraint
\begin{eqnarray}
   \phi_{2}= T(x,y) \approx 0.
   \label{7}
\end{eqnarray}
Applying the consistency conditions to this secondary constraint~(\ref{7}) yields\footnote{Where $X = \Omega x- \omega$, $Y = y - a_y$ and $\theta = 2\lambda m$.} 
%
%
%
%
\begin{eqnarray}
    \phi_{3}  =  \frac{2}{m}(X P_{x} + Y P_{y}) \approx 0,
    \label{8}
\end{eqnarray}
and by imposing the consistency condition for the constraint  $\phi_{3}$ we obtain
 \begin{equation}
    \phi_{4}= \frac{2}{m^2}[\Omega {P_x}^2 + {P_y}^2 + \theta(X^{2} + Y^{2})] \approx 0,
    \label{9}
\end{equation}
thus, a new constraint $\phi_4$ emerges, completing this step of the consistency analysis. Finally, we proceed to compute the antisymmetric matrix of constraints, collectively denoted as $\phi_{a}$ with $(a = 1, . . . , 4)$. By evaluating the Poisson brackets among all constraints, we construct the constraint matrix, which can be expressed in closed form as
%
%
%
\begin{equation}
C_{ab} \equiv \{\phi_a,\phi_b\} =
\begin{pmatrix}
0 & 0 & 0 & \{\phi_1,\phi_4\} \\
0 & 0 & \{\phi_2,\phi_3\} & \{\phi_2,\phi_4\} \\
0 & \{\phi_3,\phi_2\} & 0 & \{\phi_3,\phi_4\} \\
\{\phi_4,\phi_1\} & \{\phi_4,\phi_2\} & \{\phi_4,\phi_3\} & 0
\end{pmatrix}
\label{9a}
\end{equation}
%
%
with
\begin{small}
\begin{equation}
    \{\phi_1,\phi_4\} = \{\phi_3,\phi_2\} = -\frac{4}{m}(X^2+Y^2)
\end{equation}

\begin{equation}
    \{\phi_2,\phi_3\} = \{\phi_4,\phi_1\} = \frac{4}{m}(X^2+Y^2)
\end{equation}

\begin{equation}
    \{\phi_3,\phi_4\} = \frac{8}{m^3}\left[(\Omega^2 P^2_{x} + P^2_{y}) - \theta(\Omega X^2 + Y^2)\right]
\end{equation}

\begin{equation}
    \{\phi_2,\phi_4\} = \frac{8}{m^2}(\Omega X P_{x} + Y P_{y})
    \label{10}
\end{equation}
\end{small}
As a consequence, the Poisson bracket (PB) structure outlined above reveals that the system belongs to the class of second-class constraints, as categorized by the well-known Dirac–Bergmann classification scheme \cite{Teitelboim, hanson1964,sundermeyer1982constrained}. This identification becomes more evident when evaluating the determinant of the constraint matrix formed using the aforementioned PB relations. Given that equation (\ref{9a}) is structured as a lower-triangular matrix, it is immediately apparent that
\begin{eqnarray}
    \det C_{\text{ab}}=\bigg[\frac{4}{m}(X^{2}+Y^{2}) \bigg]^{4}. 
    \label{10a}
\end{eqnarray}
Because this determinant is non-zero, the system is confirmed to be second-class and, in particular, lacks gauge invariance. Given this second-class structure, the next step in the analysis is to construct the inverse of the constraint matrix, which is essential for defining the Dirac brackets that properly account for the imposed constraints. The inverse of equation  (\ref{9a}) is given explicitly by
%
%
%
\begin{small}
    \begin{eqnarray}
    C^{ab}=\left(
      \begin{array}{cccc}
        0 & -\rho_{1} & -\rho_{2} & \rho_{0} \\[8pt]
        \rho_{1} & 0 & - \rho_{0}& 0 \\[12pt]
        \rho_{2} & \rho_{0} & 0 & 0 \\[12pt]
       -\rho_{0} & 0 & 0 & 0
      \end{array}
    \right),
    \label{10b}
    \end{eqnarray}
\end{small}
\noindent where
\begin{eqnarray}
    \rho_{0}&=&\frac{m}{4(X^{2}+Y^{2})}, \\
    \rho_{1} &=&
\frac{-\Omega^{2}P_{x}^{2}+\theta\,\Omega\,X^{2}-P_{y}^{2}+\theta\,Y^{2}}
     {2m\,(X^{2}+Y^{2})^{2}}, \\
     \rho_{2} &=&
\frac{P_{y}\,Y+\Omega\,P_{x}\,X}
     {2\,(X^{2}+Y^{2})^{2}}.
     \label{11}
\end{eqnarray}
With the inverse of the constraint matrix at hand, we can now define the Dirac bracket, which replaces the conventional Poisson bracket when dealing with second-class systems. For any two dynamical variables $F$ and $G$ on the phase space, the Dirac bracket is defined as
%
%
\begin{equation}
\{F,G\}_{DB}=\{F,G\}-\sum^{4}_{a,b=1}\{F,\phi_{a}\}C^{ab}\{\phi_{b},G\}.
 \label{12}
\end{equation}
%
%
By substituting equation (\ref{10b}) into the definition above, one readily obtains the non-vanishing Dirac brackets between the phase space variables, which are given by

\begin{equation}
\{x, P_{x}\}_{DB} = \frac{Y^{2}}{D}
\label{12a}
\end{equation}

\begin{equation}
\{y, P_{y}\}_{DB} = \frac{X^{2}}{D}
\label{12b}
\end{equation}

\begin{equation}
\{x, P_{y}\}_{DB} = \{y, P_{x}\}_{DB} = -\frac{XY}{D}
\label{12c}
\end{equation}

\begin{equation}
\{P_{x}, P_{y}\}_{DB} = -\frac{X P_{y} - \Omega Y P_{x}}{D}
\label{12d}
\end{equation}

\begin{equation}
\{\lambda, P_{x}\}_{DB} = -\frac{P_y (P_y X - \Omega P_x Y) + X Y^2 \theta (\Omega - 1)}{m D^{2}}
\label{12e}
\end{equation}

\begin{equation}
\{\lambda, P_{y}\}_{DB} = \frac{\Omega P_x (P_y X - \Omega P_x Y) + X^2 Y \theta (\Omega - 1)}{m D^{2}}
\label{12f}
\end{equation}

\begin{equation}
\{\lambda, x\}_{DB} = -\frac{Y\left[P_y X - \Omega P_x Y \right]}{m D^{2}}
\label{12g}
\end{equation}

\begin{equation}
\{\lambda, y\}_{DB} = \frac{X\left[P_y X - \Omega P_x Y \right]}{m D^{2}}
\label{12h}
\end{equation}
%
%
%
 where $D=X^{2}+Y^{2}$. Notably, the Dirac brackets listed in \eqref{12e}–\eqref{12h} can be rewritten in a more compact form
 %
%
\begin{align}
\{\lambda, P_{x}\}_{DB} &=  \frac{Y\,\theta(\Omega - 1)}{mD} \{y, P_{x}\}_{DB} \notag \\
&\quad + \frac{P_{y}}{mD} \{P_{x}, P_{y}\}_{DB}
\end{align}
\begin{align}
\{\lambda, P_{y}\}_{DB} &= \frac{X\,\theta(1 - \Omega)}{mD} \{x, P_{y}\}_{DB} \notag \\
&\quad - \frac{\Omega P_{x}}{mD} \{P_{x}, P_{y}\}_{DB}
\end{align}
\begin{align}
\{\lambda, x\}_{DB} = \frac{Y}{m D} \{P_{x}, P_{y}\}_{DB}
\end{align}
\begin{align}
\{\lambda, y\}_{DB} = -\frac{X}{m D} \{P_{x}, P_{y}\}_{DB}
\label{13}
\end{align}
%
%
%
The structure of equations~\eqref{12a}–\eqref{12h} closely resembles that presented in \cite{barbosa2018gauge,barbosa2018brst,caires2022general}, where the authors characterize conic sections via a quadratic function with real parameters $A,...,F$. In contrast to that more general formulation, which relies on multiple parameters to characterize each conic, the present analysis adopts a streamlined approach based on a single parameter: the eccentricity \( e \). This approach significantly simplifies the analysis by allowing a direct and unified treatment of different conic types—circle (\( e = 0 \)), parabola (\( e = 1 \)), ellipse (\( 0 < e < 1 \)), and hyperbola (\( e > 1 \))—through variation of this single parameter. 

As outlined in the introduction, the Dirac brackets derived in this work extend earlier results on the dynamics of conic sections by explicitly incorporating the eccentricity as a central dynamical feature. This approach provides an additional and complementary perspective to those already reported in the literature \cite{falck1983dirac, scardicchio2002classical, nawafleh2013constrained, fung2014dirac, barbosa2018gauge, barbosa2018brst}. 

Building on this extension, the canonical quantization procedure proceeds by replacing Dirac brackets with the corresponding quantum commutators\footnote{$ [\hat{q}_{i},\hat{P}^{j}]=i\hbar\{q_{i},P^{j}\}_{DB}$.} 
thereby ensuring a consistent implementation of the system’s constraints at the quantum level \cite{barbosa2018gauge}. Ambiguities arising from operator ordering are naturally resolved by imposing Hermiticity on the quantum operators, as illustrated in the case of a particle constrained to move on a circle \cite{scardicchio2002classical}.

Once all constraints have been identified and the Dirac brackets derived, we proceed to simplify the canonical Hamiltonian~\eqref{4e} by eliminating those constraints that appear explicitly. This step allows us to compute the equations of motion for each dynamical variable within the reduced framework. As a result of this procedure, the reduced Hamiltonian takes the form

%
\begin{equation}
    H_c \;=\; \frac{1}{2m}\bigl(P_x^{2} + P_y^{2}\bigr).
    \label{14}
\end{equation}
%
%
%
%
%
Thus, the Hamiltonian~\eqref{14} describing the constrained particle retains a structure closely resembling that of the unconstrained case, resulting in analogous equations of motion in both settings. Accordingly, the classical equations of motion can be expressed in terms of Dirac brackets as follows
\begin{equation}
\dot{x} = -\frac{{Y}\,{\left(X\,P_y -Y\,P_x\right)}}{m\,{D}},
\label{15}
\end{equation}
\begin{equation}
\dot{y} = \frac{{X}\,{\left(X\,P_y -Y\,P_x\right)}}{m\,{D}},
\label{16}
\end{equation}
\begin{equation}
\dot{P}_x = -\frac{{P_y}\,{\left(X\,P_y -Y\,\Omega\,P_x\right)}}{m\,{D}},
\label{17}
\end{equation}
\begin{equation}
\dot{P}_y = \frac{{P_x}\,{\left(X\,P_y -Y\,\Omega\,P_x\right)}}{m\,{D}}.
\label{18}
\end{equation}
In the following section, we extend the model to its relativistic formulation and examine the resulting modifications to the constraint structure, Dirac brackets, and equations of motion.
%
%
\section{Relativistic particle constrained on 
a non-degenerate conic section}
%
%
%
We now extend our analysis to the relativistic regime of the system previously studied in its non-relativistic form. To this end, we replace the Lagrangian in equation~(\ref{2}) with its relativistic counterpart
\begin{small}
\begin{eqnarray}
L=-c^2 \,m\,\sqrt{1-c^{-2}({\dot{x} }^2 +{\dot{y} }^2) }+\lambda\,T(x,y),
\label{18a}
\end{eqnarray}
\end{small}

with m and c denoting, respectively, the particle mass and universal speed of light. \footnote{As in the previous section, the multiplier \(\lambda\) will be considered as a dynamical variable of the system.}


To determine the system’s singularity structure, we compute the Hessian matrix of second derivatives of the Lagrangian with respect to the generalized velocities 
\begin{small}
\begin{equation}
W=\frac{m}{c^2 \,{{\left(1- c^{-2}({\dot{x} }^2 +{\dot{y} }^2) \right)}}^{3/2}} \left(
\begin{array}{cccc}
c^2 - \dot{y}^2 & \dot{x}\,\dot{y} & 0 & \\
\dot{x}\,\dot{y} & c^2 - \dot{x}^2 & 0 \\
0 & 0 & 0 \\
\end{array}
\label{18b}
\right).
\end{equation}
\end{small}
%

As expected, the Hessian is degenerate, confirming that the system is singular and subject to constraints. To formulate the corresponding Hamiltonian description, we begin by calculating the canonical momenta $(P_{x},P_{y},P_{\lambda})$ associated with each  coordinate $(x,y,\lambda)$
%
%
%
%
\begin{equation}
\begin{aligned}
P_{x} &= \frac{m\,\dot{x}}{\sqrt{1 - \frac{\dot{x}^2 + \dot{y}^2}{c^2}}}, \quad
P_{y} = \frac{m\,\dot{y}}{\sqrt{1 - \frac{\dot{x}^2 + \dot{y}^2}{c^2}}}, \quad
P_{\lambda} = 0
\end{aligned}
\label{19}
\end{equation}
from which we identify the primary constraint
%
%
%
\begin{equation}
\tilde{\phi}_1 = P_{\lambda} \approx 0.
\label{19a}
\end{equation}
%



Following the approach presented in \cite{oliveira2024relativistic}, we begin by introducing the canonical momenta $(P_{x},P_{y},P_{\lambda})$
 associated with the configuration variables. This allows us to construct the canonical Hamiltonian, which takes the following form
\begin{equation}
    H_c =H_{0}-\lambda\,T(x,y),
    \label{20}
\end{equation}
where
\begin{eqnarray}
    H_0 = c \,\sqrt{c^2\,m^2+ ({P_x }^2 +{P_y }^2) },
    \label{20a}
\end{eqnarray}
represents the free relativistic Hamiltonian.



Using the primary constraint~(\ref{19a}), we derive the following secondary constraint by ensuring consistency of the constraint evolution under the total Hamiltonian\footnote{Due to the length of the ensuing expressions, we introduce another redefinition of variables: \(\Pi = X P_y - Y P_x\)
}.
%
%
\begin{equation}
\tilde{\phi}_{2} = T(x,y) \approx 0
\label{21}
\end{equation}
\begin{equation}
\tilde{\phi}_{3} = \frac{2 c^{2}}{H_{0}} (X P_{x} + Y P_{y}) \approx 0
\label{22}
\end{equation}
\begin{align}
\tilde{\phi}_{4} &= \frac{2 c^{4}}{H_0^{2}} \bigg[
(\Omega P_{x}^{2} + P_{y}^{2}) 
+ \frac{c^{2} m \theta}{H_0}D + \frac{\theta}{m H_0}\Pi^{2}
\bigg] \approx 0
\label{23}
\end{align}
%
%
To simplify the structure of the constraint algebra, we first remove the overall prefactor from~\eqref{22}, yielding the equivalent constraint
\begin{eqnarray}
    \phi_{3}  =  X P_{x} + Y P_{y} \approx 0. 
    \label{24}
\end{eqnarray}
Evolving this reduced constraint under the total Hamiltonian leads to a fourth constraint in simplified form
\begin{equation}
\phi_{4} = \frac{c^{2}}{H_{0}} \left( \Omega P_{x}^{2} + P_{y}^{2} \right) + \frac{\theta}{m}D \approx 0.  \label{24a}
\end{equation}
Although constraints~(\ref{22}) and~(\ref{23}) differ in form from their simplified versions~(\ref{24}) and~(\ref{24a}), they are entirely equivalent. Notably, while the simplified constraints yield a Poisson bracket matrix that is structurally distinct from that produced by the extended forms, both sets lead to identical Dirac brackets. The advantage of the simplified formulation lies in its ability to eliminate redundant terms automatically. In contrast, the extended expressions require explicit substitution of the constraint definitions to achieve the same simplification. Simplifying the expressions in this way can mischaracterize the system, so we generally recommend using the extended form of the constraints. In this study, we examined both the reduced and extended expressions and obtained identical results, showing that the simplification is acceptable in this case.
 Due to the unwieldy size and complexity of the Poisson matrix associated with the extended constraints, we proceed using the simplified set. For a detailed derivation using the full extended constraints, readers may refer to Appendix~\ref{AppendixB}.

Finally, we compute the antisymmetric matrix of Poisson brackets among the constraints~(\ref{19a}), (\ref{21}), (\ref{24}), and (\ref{24a})
%
\begin{equation}
C_{ab} = \{\phi_a, \phi_b\} =
2\begin{pmatrix}
0 & 0 & 0 & -\alpha_1 \\
0 & 0 & \alpha_1 & \alpha_2 \\
0 & -\alpha_1 & 0 & \alpha_3 \\
\alpha_1 & -\alpha_2 & -\alpha_3 & 0
\end{pmatrix},
\label{24b}
\end{equation}
where
\begin{equation}
\alpha_1 = X^2 + Y^2
\label{25a}
\end{equation}
\begin{align}
\alpha_2 &= \frac{2 c^2}{H_0} \bigg[ 
\Omega X P_x + Y P_y
\bigg]
\label{25b}
\end{align}
\begin{align}
\alpha_3 &= \frac{c^2}{H_0} \left[
\Omega^2 P_x^2 + P_y^2 
- \frac{c^2}{2 H_0^2} (\Omega P_x^2 + P_y^2)^2 
\right] \notag \\
&\quad - \frac{1}{m} \theta (\Omega X^2 + Y^2)
\label{25c}
\end{align}
and its inverse
\begin{small}
\begin{eqnarray}
C^{ab}= \frac{1}{2}\left(
  \begin{array}{cccc}
     0 & -\beta_{1}  & \beta_{2} & \beta_{3} \\[4pt]
    \beta_{1} & 0 & -\beta_{3} & 0 \\[8pt]
     -\beta_{2} & \beta_3 & 0 & 0 \\[12pt]
     -\beta_3 & 0 & 0 & 0
  \end{array}
\right),
\label{24c}
\end{eqnarray}
\end{small}
whit
\begin{align}
\beta_1 &= \frac{c^4}{2 H_0^3 D^2} \bigg[
\frac{2 H_0^3}{m c^4} \theta (\Omega X^2 + Y^2) \notag \\
&\quad - \left( \frac{2 H_0^2}{c^2}(\Omega^2 P_x^2 + P_y^2) 
- (\Omega P_x^2 + P_y^2)^2 \right)
\bigg]
\label{26a}
\end{align}
\begin{align}
\beta_2 &= -\frac{2\,c^2}{H_0 D^2} \bigg[
 \notag  (\Omega X P_x + Y P_y)
\bigg]
\label{26b}
\end{align}
\begin{equation}
\beta_3 = \frac{1}{D}
\label{26c}
\end{equation}
Starting from \eqref{12}, one can compute the Dirac brackets for any pair of phase space variables as follows
\begin{eqnarray}
\{x, P_{x}\}_{DB}&=&\frac{Y^{2}}{D}, \label{25aa} \\
\{y, P_{y}\}_{DB}&=& \frac{X^{2}}{D} , \label{25bb} \\
\{x, P_{y}\}_{DB} &=&\{y, P_{x}\}_{DB} = -\frac{XY}{D} , \label{25cc} \\
\{P_{x}, P_{y}\}_{DB}&=& -\frac{X P_{y}-\Omega Y P_{x}}{D}.
\label{25d} 
\end{eqnarray}
The Dirac brackets between the physical coordinates \eqref{25aa}-\eqref{25d} reveal that the fundamental algebraic structure of the system, characterized as a second-class model, remains intact when relativistic effects are incorporated. However, the brackets involving the parameter $\lambda$ exhibit significant modifications\footnote{For completeness, we include in Appendix~\ref{AppendixB} the full calculation of all non-zero Dirac brackets that involve the constraint coordinate $\lambda$, completing our description of the system's algebraic structure.}. These findings closely align with those reported in \cite{oliveira2024relativistic} and \cite{caires2022general}, where both relativistic and non-relativistic cases are studied, explicitly accounting for constraints and the presence of an open potential.

Moreover, the Dirac brackets derived here extend and generalize previous results concerning specific conic sections, explicitly formulated in terms of the eccentricity parameter $e$. This provides a distinct perspective consistent with earlier works \cite{caires2022general, barbosa2018gauge, fung2014dirac, scardicchio2002classical, thibes2021brst, nawafleh2013constrained, najafizade2022dirac}.


In order to calculate the motion equations we use the reduce Hamiltonian
\begin{equation}
    H_c =H_{0} = c \,\sqrt{m^2\,c^2+{P_x }^2 +{P_y }^2},
    \label{26}
\end{equation}
obtaining the motion equations
\begin{eqnarray}
    \dot{x} &=& -\frac{c^2}{H_0}\,\left[ \frac{{Y}\,{\Pi}}{{D}} \right],
    \label{27}\\
    \dot{y} &=& \frac{c^2}{H_0}\,\left[\frac{{X}\,{\Pi}}{{D}} \right], \label{28}\\
    \dot{P}_x &=& -\frac{c^2}{H_0}\,\left[\frac{{P_y}\,{\left(X\,P_y -Y\,\Omega\,P_x\right)}}{{D}} \right], \label{29}\\
    \dot{P}_y &=& \frac{c^2}{H_0}\,\left[\frac{{P_x}\,{\left(X\,P_y -Y\,\Omega\,P_x\right)}}{{D}} \right].\label{30}
\end{eqnarray}
%
%
%
%
Analyzing the system of equations \eqref{27}-\eqref{30}, we observe that in the low-velocity limit ($v \ll c$), the expression $\dfrac{c^{2}}{H_{0}}$ reduces to
\begin{equation}
\lim_{\substack{\text{low} \\ \text{velocities}}} \frac{c^{2}}{H_{0}} = \dfrac{1}{m}.
\label{31}
\end{equation}
In this limit, the equations reduce to those derived in the previous section, as given by \eqref{15}-\eqref{18}. 

\section{Hamilton-Jacobi Analysis}


This section provides a concise overview of the HJ method, following the treatments presented in \cite{guler1992canonical, Bertin, Bertin2, romero2025singular}. From the conjugate momenta presented in \eqref{4b}, we deduce the existence of a primary constraint, expressed by \eqref{4c}.
Furthermore, the canonical variables satisfy the following Poisson bracket relations
\begin{equation}
\{x, P_{x}\} = 1, \quad \{y, P_{y}\} = 1, \quad \{\lambda, P_{\lambda}\} = 1
\label{32a}
\end{equation}
The canonical formulation \cite{guler1992canonical,Na} leads to the corresponding set of HJPDEs\footnote{Where $S$ is a Hamilton's principal function and $P_0 = \partial_t S$.}, given by
\begin{eqnarray}
    H^{'}_0 &=& P_0 + H_c = 0, \label{33}\\
    H^{'}_{1}&=& P_{\lambda}=0. \label{34}
\end{eqnarray}
To explore whether the system is integrable, we focus on how a generic dynamical function evolves over time, working within the framework of Carathéodory’s formalism \cite{guler1992canonical, Nawafleh1}. In this setting, the key object of analysis is the fundamental differential, which takes the following form
\begin{equation}\label{37}
    dF = \left\{F, H^{'}_0\right\}dt +  \left\{F, H^{'}_1\right\}d\lambda .
\end{equation}
Since $\{H^{'}_{1},H^{1}_{1}\}=0$, the consistency condition for $H^{'}_{1}$
\begin{eqnarray}
dH^{'}_{1} &=& \{H^{'}_{1},H^{'}_{0}\}dt+ \{H^{'}_{1},H^{'}_{1}\}d\lambda= 0,
\label{38}
\end{eqnarray}
leads to the emergence of a secondary constraint given by
\begin{equation}
    H^{'}_{2}= T(x,y)=0.
    \label{39}
\end{equation}
Similarly, by evolving this constraint and the subsequent ones through equation \eqref{37}, we recover the same final set of constraints previously identified in Section 3: $\left\{\phi_1 \rightarrow H^{'}_1, \phi_2 \rightarrow H^{'}_2, \phi_3 \rightarrow H^{'}_3, \phi_4 \rightarrow H^{'}_4 \right\}$, to which we associate the following independent parameters respectively: $\left\{\lambda, \alpha, \beta, \gamma \right\}$. 
Next, we proceed to verify the integrability of this set through the extended fundamental differential
\begin{align}
dF &= \left\{F,  H^{'}_0\right\}dt 
+ \left\{F,  H^{'}_{1}\right\}d\lambda 
+ \left\{F,  H^{'}_{2}\right\}d\alpha \notag \\
&\quad + \left\{F,  H^{'}_{3}\right\}d\beta 
+ \left\{F,  H^{'}_{4}\right\}d\gamma
\label{40}
\end{align}
Thus, the time evolution of the constraints is given by
\begin{equation}
dH^{'}_{1} = \left\{ H^{'}_1,  H^{'}_0\right\} dt + \left\{ H^{'}_1,  H^{'}_{4}\right\} d\gamma
\label{40a}
\end{equation}
\begin{align}
dH^{'}_{2} &= \left\{H^{'}_{2},  H^{'}_0\right\} dt 
+ \left\{H^{'}_{2},  H^{'}_{3}\right\} d\beta \notag \\
&\quad + \left\{H^{'}_{2},  H^{'}_{4}\right\} d\gamma
\label{40b}
\end{align}
\begin{align}
dH^{'}_{3} &= \left\{H^{'}_{3},  H^{'}_0\right\} dt 
+ \left\{H^{'}_{3},  H^{'}_{2}\right\} d\alpha \notag \\
&\quad + \left\{H^{'}_{3},  H^{'}_{4}\right\} d\gamma
\label{40c}
\end{align}
\begin{align}
dH^{'}_{4} &= \left\{H^{'}_{4},  H^{'}_0\right\} dt 
+ \left\{H^{'}_{4},  H^{'}_{1}\right\} d\lambda \notag \\
&\quad + \left\{H^{'}_{4},  H^{'}_{2}\right\} d\alpha 
+ \left\{H^{'}_{4},  H^{'}_{3}\right\} d\beta
\label{41}
\end{align}
The set of HJPDEs obtained above is not in involution, as can be verified through the computation of their Poisson bracket \eqref{10}. A direct computation reveals that the system governed by \eqref{40a}-\eqref{41} can be written as
\begin{equation}
dH^{'}_{1} =  T(x,y) \, dt + \sigma_1\, d\gamma
\label{42}
\end{equation}
\begin{equation}
dH^{'}_{2} = \frac{2}{m} (X P_x + Y P_y)\, dt - \sigma_1\, d\beta + \sigma_3\, d\gamma
\label{43}
\end{equation}
\begin{equation}
dH^{'}_3 = \frac{2}{m^2} \left[ \Omega P_x^2 + P_y^2 + \theta D \right] dt + \sigma_1\, d\alpha + \sigma_2\, d\gamma
\label{44}
\end{equation}
\begin{equation}
dH^{'}_4 = 2 \lambda \sigma_3\, dt - \sigma_1\, d\lambda - \sigma_3\, d\alpha - \sigma_2\, d\beta
\label{45}
\end{equation}
 where $\sigma_{i}=(\sigma_{1},\sigma_{2},\sigma_{3})$,  are defined as follows 
\begin{align}
\sigma_1 &= -\{H^{'}_{1}, H^{'}_{4}\} = \{H^{'}_{2}, H^{'}_{3}\} 
= -\{H^{'}_{3}, H^{'}_{2}\} \notag \\
&= \{H^{'}_{4}, H^{'}_{1}\} 
= \frac{4}{m}\,D
\label{46}
\end{align}
\begin{equation}
\sigma_2 = \{H^{'}_{3}, H^{'}_{4}\}
= \frac{8}{m^3}\left[(\Omega^2 P_x^2 + P_y^2) - \theta(\Omega X^2 + Y^2)\right]
\label{47}
\end{equation}
\begin{equation}
\sigma_3 = \{H^{'}_{2}, H^{'}_{4}\}
= \frac{8}{m^2}(\Omega X P_x + Y P_y)
\label{48}
\end{equation}
The evolution of the constraints clearly reveals that several of their Poisson brackets are non-vanishing, confirming that the set is non-involutive. Consequently, it becomes necessary to construct the generalized brackets to achieve a consistent dynamical framework.

We start by calculating the Poisson bracket matrix associated with the full set of constraints, which we define as $M_{\mu\nu}=\{H^{'}_{\mu}, H^{'}_{\nu}\}$, with $\mu,\nu=1,2,3,4.$ We found it notable that this matrix $M$ exactly coincides with the one obtained earlier in equation (\ref{9a}), since both are constructed from the identical set of constraints previously identified using Dirac’s formalism. Moreover, the fundamental brackets in the HJ framework exhibit the same algebraic structure as those introduced in Dirac’s approach, as shown in equation (\ref{12}). We chose this framework because the Generalized Brackets naturally take the same form as Dirac Brackets \eqref{12a}-\eqref{12h}. From this, the equations of motion derived using these Generalized Brackets are 
\begin{equation}
\dot{x} = \frac{Y \Sigma}{m D^2}
\label{49}
\end{equation}
\begin{equation}
\dot{y} = -\frac{X \Sigma}{m D^2}
\label{50}
\end{equation}
\begin{equation}
\dot{P}_x = \frac{\Gamma_1}{m D^2}
\label{51}
\end{equation}
\begin{equation}
\dot{P}_y = \frac{\Gamma_2}{m D^2}
\label{52}
\end{equation}
where
%
%
%
%
%
\begin{equation}
\Sigma \;=\;
      -\Pi \,D
      -\bigl(P_y\,X - \Omega\,P_x\,Y\bigr)\,
       H^{'}_{2},
       \label{53}
\end{equation}
\begin{align}
\Gamma_1 \;=\;&-P_y \,{\left(P_y \,X-\Omega \,P_x \,Y\right)}D \notag \\
&-{\left(X\,{P_y }^2 -\Omega \,P_x \,P_y \,Y+X\,\theta \,{\left(\Omega -1\right)}\,Y^2 \right)}H^{'}_{2},
\label{54}
\end{align}
\begin{align}
\Gamma_2 \;=\;&P_x \,{\left(P_y \,X-\Omega \,P_x \,Y\right)}D \notag \\
&+ {\left(X^2 \,Y\,\theta \,{\left(\Omega -1\right)}-\Omega^2 \,{P_x }^2 \,Y+\Omega \,P_x \,P_y \,X\right)}\,H^{'}_{2}.
\label{55}
\end{align}
By substituting the definition of the constraint \(H^{'}_2\) into equations \eqref{53}, \eqref{54}, and \eqref{55}, we obtain the following simplified expressions
\begin{eqnarray}
    \Sigma &=&
      -\Pi\,D, \label{56} \\
      \Gamma_1 &=&
      -P_y \,{\left(P_y \,X-\Omega \,P_x \,Y\right)}D, \label{57} \\
      \Gamma_2 &=&
P_x \,{\left(P_y \,X-\Omega \,P_x \,Y\right)}D. \label{58}
\end{eqnarray}
When these reduced forms \eqref{56}–\eqref{58} are inserted into the equations of motion \eqref{49}–\eqref{52}, they lead to the same dynamical equations previously obtained via the Dirac formalism, as shown in Eqs.~\eqref{15}–\eqref{18}.

\section{Relativistic version via HJ}

We now turn to the relativistic analysis of a particle constrained to move along a conic section, using the HJ formalism. The system’s behavior is then determined by the Lagrangian showed on \eqref{18a}, therefore, we obtain the same conjugate momenta \eqref{19} and canonical Hamiltonian \eqref{20}.

In this way, we can count with the following set of HJPDE's 
\begin{eqnarray}
    H^{'}_0 &=& P_0 + H_c = 0, \label{59}\\
    H^{'}_1 &=& P_\lambda = 0, \label{60}
\end{eqnarray}
and the Fundamental Differential being
\begin{equation}\label{61}
    dF= \left\{F,H^{'}_0 \right\}dt + \left\{F, H^{'}_1 \right\} d\lambda,
\end{equation}
which lead us to get the subsequent constraints:
%
\begin{equation}
H^{'}_{2} = T(x,y)
\label{62}
\end{equation}
\begin{equation}
H^{'}_{3} = \frac{2 c^2}{H_0} (X P_x + Y P_y)
\label{63}
\end{equation}
\begin{align}
H^{'}_{4} &= \frac{2 c^4}{H_0^2} \bigg[
\Omega P_x^2 + P_y^2 
+ \frac{c^2 m \theta}{H_0}D + \frac{\theta}{m H_0}\Pi^2
\bigg]
\label{64}
\end{align}
%
%
The fact that both  $\left\{ H^{'}_4, H^{'}_0 \right\}$ and $\left\{ H^{'}_4, H^{'}_1 \right\}$ are non-zero implies that no further constraints are generated, confirming the non-involutive nature of the system.

We now proceed to verify the integrability conditions associated with the system’s constraints. This verification is performed using the extended fundamental differential 
%
%
\begin{align}
dF &= \left\{F, H^{'}_0 \right\} dt 
+ \left\{F, H^{'}_1 \right\} d\lambda 
+ \left\{F, H^{'}_2 \right\} d\alpha \notag \\
&\quad + \left\{F, H^{'}_3 \right\} d\beta 
+ \left\{F, H^{'}_4 \right\} d\gamma
\label{65}
\end{align}

Here, the variables  $\left(\alpha, \beta, \gamma\right)$
are naturally associated with the new constraints $(H^{'}_2, H^{'}_3, H^{'}_4)$, respectively.


Upon analyzing the complete set of HJPDEs—specifically  Eq.\eqref{65}, it becomes evident that the system is non-involutive. This observation signals the presence of second-class constraints and underscores the need to introduce a modified bracket structure to ensure the consistent evolution of the system’s dynamics.




To address this issue, we formulate the equations of motion using the generalized bracket—analogous to the Dirac bracket introduced in Eq.~\eqref{12}—which is defined as
\begin{equation}
\dot{F} = \left\{F, \phi_0 \right\}^{*} = \left\{F, \phi_0 \right\} 
- \left\{F, \phi_{\nu} \right\} \left(M^{-1}\right)^{\nu \mu} \left\{ \phi_{\mu}, \phi_0 \right\}
\label{66}
\end{equation}

here $F$ represents any dynamical variable, and $M^{-1}$
denotes the inverse of the matrix composed of Poisson brackets among the non-involutive constraints. 
This generalized bracket structure ensures that the system’s evolution remains consistent with all imposed constraints. By applying Eq.~\eqref{66} to compute the time evolution of the relevant variables, while simultaneously enforcing $H^{'}_{i}=0$, one obtains the following set of equations \footnote{The expressions associated with the variable $\lambda$ can be found in Appendix~\ref{AppendixB}.}
%
%
%
\begin{eqnarray}
    \dot{x} &=& \frac{\,H_0 \,\Pi^2\, Y}{ \Pi^2 +  m^2  c^2D} \label{66a}\\
     \dot{y} &=& \frac{ \,H_0\,\Pi^2\, X }{ \Pi^2 + m^2 c^2 D} \label{66b}\\
     \dot{P}_x &=& \frac{H_0 \left( \Omega Y P_x - X P_y \right) P_y  }{\Pi + m^2 c^2 D}\label{66d} \\
     \dot{P}_y &=& \frac{H_0 \left( X P_y - \Omega Y P_x \right) P_x  }{ \Pi  + m^2 c^2  D}\label{66f}\\
\end{eqnarray}
%
%
Now the GB dynamics:
\begin{equation}
\left\{ x, P_x \right\}^{*} = -\frac{(\Pi\,P_x  - m^2 c^2 Y) Y}
{\Pi^2 + m^2 c^2 D}
\label{67a}
\end{equation}
\begin{equation}
\left\{ y, P_y \right\}^{*} = \frac{(\Pi\,P_y + m^2 c^2 X) X}
{ \Pi^2 + m^2 c^2 D}
\label{67b}
\end{equation}
\begin{equation}
\left\{ x, P_y \right\}^{*} = -\frac{(\Pi \, P_y + m^2 c^2 X) Y}
{\left( \Pi \right)^2 + m^2 c^2 D}
\label{67c}
\end{equation}
\begin{equation}
\left\{ y, P_x \right\}^{*} = -\frac{(\Pi\,P_x - m^2 c^2 Y) X}
{\Pi^2 + m^2 c^2 D}
\label{67d}
\end{equation}
\begin{equation}
\left\{ P_x, P_y \right\}^{*} = 
\frac{(m^2 c^2 + P_x^2 + P_y^2)(\Omega Y P_x - X P_y)}
{\Pi^2 + m^2 c^2 D}
\label{67e}
\end{equation}
From the equations of motion and the GB dynamics, it is easy to denote that applying the non-relativistic limit $c \rightarrow \infty $, which implies $\tfrac{H_0}{c^2} \rightarrow m$, one recover the equations of motion \eqref{49} - \eqref{52} from the classic case.

Furthermore, this system initially counts with six phase space dynamical variables, but the presence of four non-involutive (second-class) constraints is the reason for the number of degrees of freedom at configuration space level that one needs for describing being one.

\section{Conclusion}

This work rigorously demonstrates the application of Dirac's constraint formalism, emphasizing the critical importance of a thorough and complete algorithmic execution at each stage. As highlighted, simplifying steps or omitting key verifications, while seemingly innocuous in certain cases, can lead to significant ambiguities in the analysis of constrained systems \cite{brown2022singular}. Our study reinforces that strict adherence to the Dirac-Bergmann procedure is crucial for robust and unambiguous results.

A central achievement of this research is the comprehensive analysis of a particle's motion on a conic section, investigated in both classical (non-relativistic) and relativistic regimes. In the non-relativistic framework, the Dirac-Bergmann algorithm precisely identified four second-class constraints. This identification allowed for the explicit computation of Dirac brackets, which inherently depend on the eccentricity parameter e, and the subsequent determination of the physical degrees of freedom. This methodology yielded a unified treatment for all non-degenerate conic trajectories—circular ($e=0$), elliptical ($0<e<1$), parabolic ($e=1$), and hyperbolic ($e>1$)—within a single, consistent formal framework. Crucially, extending this analysis to the relativistic domain revealed an analogous constraint structure, preserving the second-class nature of the system and enabling successful canonical quantization via Dirac brackets despite increased algebraic complexity.

Complementing the Dirac approach, the Hamilton-Jacobi (HJ) formalism was applied to the same system. Initially, this investigation uncovered a set of non-involutive constraints. However, the introduction of generalized brackets successfully restored the system's integrability, leading to the accurate derivation of the equations of motion. The consistency of our findings was affirmed by a direct comparison, showing that the equations of motion derived from the Hamilton-Jacobi method precisely coincide with those obtained through Dirac's formalism.

This combined and extended analysis significantly deepens the understanding of constrained dynamics under both classical and relativistic conditions. Furthermore, it clearly demonstrates the complementary strengths of the Dirac and Hamilton-Jacobi formalisms in addressing singular systems. While Dirac's approach systematically classifies and manages constraints to define the phase space structure and dynamics, the HJ method offers a powerful alternative for solving equations of motion, particularly when augmented with generalized brackets for non-involutive constraints. This unified framework not only elucidates the geometric structure of conic trajectories but also establishes a robust foundation for future investigations into classical and quantum constrained systems.

\bibliography{sn-bibliography}

\newpage

\appendix
\section{Appendix: Derivation and Elements of the Conic Section Equation}
\label{AppendixA}

This appendix presents a detailed derivation of \eqref{3}, which describes a general conic section. That equation follows directly from the geometric definition of a conic as the locus of points \(P=(x,y)\) in the plane for which the ratio of the distance to a fixed point (the focus) and the distance to a fixed line (the directrix) equals a positive constant (the eccentricity).

The conic section equation is:
\begin{equation}
    (y - a_{y})^{2} + (1 - e^{2})\,x^{2} 
    + 2\,(u\,e^{2} - a_{x})\,x 
    + a_{x}^{2} - u^{2}\,e^{2} 
    = 0.
    \label{A1}
\end{equation}
Below, we explain how each term arises and how the parameters are defined. To derive \eqref{A1}, we begin with the following elements:
\begin{itemize}
    \item \textbf{Focus (\(F\))}: A fixed point \(F = (a_{x},\,a_{y})\) in the Euclidean plane \(\mathbb{E}^{2}\). The coordinates \((a_{x},\,a_{y})\) are constants that specify the focus location for a given conic.
    \item \textbf{Directrix (\(\delta\))}: A fixed line \(\delta\) given by \(x = u\), where \(u \neq a_{x}\). Requiring \(u \neq a_{x}\) ensures the focus does not lie on the directrix, avoiding degenerate cases.
    \item \textbf{Generic Point (\(P\))}: A variable point \(P = (x,y)\) in \(\mathbb{E}^{2}\) that must satisfy the conic definition.
    \item \textbf{Eccentricity (\(e\))}: A positive constant \(e > 0\) defined by
\begin{align}
  d\bigl(P,F\bigr) &= e\,d\bigl(P,\delta\bigr), \label{eq:def_conic}\\[4pt]
  d\bigl(P,F\bigr) &= \sqrt{(x - a_x)^2 + (y - a_y)^2}, \\[4pt]
  d\bigl(P,\delta\bigr) &= \lvert x - u \rvert.
\end{align}
\end{itemize}
By definition of a conic, we set
\begin{equation}
    \sqrt{(x - a_{x})^{2} + (y - a_{y})^{2}}
    \;=\; 
    e\,\lvert x - u\rvert\,.
    \label{A2}
\end{equation}
Starting from \eqref{A2}, we square both sides:
\begin{equation}
    (x - a_{x})^{2} + (y - a_{y})^{2}
    \;=\; 
    e^{2}\,(x - u)^{2}.
    \label{A3}
\end{equation} 
Expanding each term:
\begin{align*}
    (y - a_{y})^{2} &\text{ remains as is},\\
    (x - a_{x})^{2} &= x^{2} - 2\,a_{x}\,x + a_{x}^{2},\\
    e^{2}\,(x - u)^{2} &= e^{2}\bigl(x^{2} - 2\,u\,x + u^{2}\bigr) \\
    &= e^{2}\,x^{2} - 2\,u\,e^{2}\,x + e^{2}\,u^{2}.
\end{align*}
Substitute into \eqref{A3}:
\begin{equation*}
        x^{2} - 2\,a_{x}\,x + a_{x}^{2}
    + (y - a_{y})^{2}
    = 
    e^{2}\,x^{2} - 2\,u\,e^{2}\,x + e^{2}\,u^{2}.
\end{equation*}
Rearrange all terms to one side:
\begin{equation}
        (1 - e^{2})\,x^{2}
    + 2\,(u\,e^{2} - a_{x})\,x 
    + a_{x}^{2} - e^{2}\,u^{2} 
    + (y - a_{y})^{2} 
    = 0.
\end{equation}
This is precisely Equation \eqref{A1}. Each term in Equation~\eqref{A1} possesses a clear geometric origin:
\begin{itemize}
    \item \((y - a_{y})^{2}\): Arises from the vertical component of \(\sqrt{(x - a_{x})^{2} + (y - a_{y})^{2}}\) after squaring.
    \item \((1 - e^{2})\,x^{2}\): The coefficient \(1 - e^{2}\) multiplies \(x^{2}\), combining \(x^{2}\) from \((x - a_{x})^{2}\) with \(-e^{2}\,x^{2}\) from \(e^{2}(x - u)^{2}\).
    \item \(2\,(u\,e^{2} - a_{x})\,x\): The linear term in \(x\) results from \(-2\,a_{x}\,x\) (focus part) and \(+2\,u\,e^{2}\,x\) (directrix part times \(-1\)).
    \item \(a_{x}^{2} - e^{2}\,u^{2}\): These constants come from \(a_{x}^{2}\) in \((x - a_{x})^{2}\) and \(-e^{2}\,u^{2}\) in \(e^{2}(x - u)^{2}\).
\end{itemize}

The sign of \(1 - e^{2}\) determines the conic’s shape:
\begin{itemize}
    \item \emph{Ellipse} (\(0 < e < 1\)): \(1 - e^{2} > 0\), so the \(x^{2}\)-term is positive.
    \item \emph{Parabola} (\(e = 1\)): \(1 - e^{2} = 0\), eliminating the \(x^{2}\)-term.
    \item \emph{Hyperbola} (\(e > 1\)): \(1 - e^{2} < 0\), making the \(x^{2}\)-term negative.
\end{itemize}
Thus, a single parameter \(e\) unifies all standard conic sections in one equation.
Observe that when \(e = 0\), Equation~\eqref{A1} reduces to
\begin{equation*}
       (y - a_{y})^{2} + (x - a_{x})^{2} = 0,
\end{equation*}
which represents a \emph{degenerate, imaginary conic}. Equivalently, it factors as
\begin{equation*}
        \bigl(y - a_{y} + i\,(x - a_{x})\bigr)\,\bigl(y - a_{y} - i\,(x - a_{x})\bigr) \;=\; 0.
\end{equation*}
If instead one fixes \(e^{2}\,(u - a_{x})^{2} = r^{2} \neq 0\) and then takes the limit \(e \to 0\), Equation~\eqref{A1} becomes
\begin{equation*}
        (x - a_{x})^{2} \;+\; (y - a_{y})^{2} \;=\; r^{2},
\end{equation*}
which is the equation of a real circle of radius \(r\). In other words, a circle can be viewed as the limiting case of a family of ellipses as \(e \to 0\).



\section{Appendix: Unsimplified Constraints in the Relativistic System}
\label{AppendixB}

This appendix presents the results obtained for the system’s dynamics when the constraints are kept in their full, unsimplified form.  
All the expressions were generated with the \textsc{Matlab} program developed for this work (as were the other results reported herein), so the equations shown below are the raw outputs produced by that code.

We begin by writing the constraints in their extended form
\begin{small}
\begin{equation}
\tilde{\phi}_{1} = P_\lambda
\end{equation}
\begin{equation}
\tilde{\phi}_{2} = T(x,y) \approx 0
\label{21b}
\end{equation}
\begin{equation}
\tilde{\phi}_{3} = \frac{2 c^{2}}{H_{0}} (X P_{x} + Y P_{y}) \approx 0
\label{22b}
\end{equation}
\begin{align}
\tilde{\phi}_{4} =\; & \frac{2 c^{4}}{H^{2}_{0}} \Bigg[
    (\Omega P_{x}^{2} + P_{y}^{2})  + \frac{c^{2}m\theta}{H_{0}}D 
    + \frac{\theta}{mH_0}\Pi^2
\Bigg] \approx 0
\label{B1}
\end{align}

\end{small}

Using these constraints, we obtain the following Dirac Brackets:
\begin{equation}
\{x, P_{x}\}_{DB} = \frac{Y(Y\,{P_x}^2 - X\,P_x\,P_y + Y\,m^2\,c^2)}{m^2\,c^2(X^{2} + Y^{2}) + (X\,P_y - Y\,P_x)^2}
\end{equation}
\begin{equation}
\{y, P_{y}\}_{DB} = \frac{X(X\,{P_y}^2 - Y\,P_x\,P_y + X\,m^2\,c^2)}{m^2\,c^2(X^{2} + Y^{2}) + (X\,P_y - Y\,P_x)^2}
\end{equation}
\begin{equation}
\{x, P_{y}\}_{DB} = -\frac{Y(X\,{P_y}^2 - Y\,P_x\,P_y + X\,m^2\,c^2)}{m^2\,c^2(X^{2} + Y^{2}) + (X\,P_y - Y\,P_x)^2}
\end{equation}
\begin{equation}
\{y, P_{x}\}_{DB} = -\frac{X(Y\,{P_x}^2 - X\,P_x\,P_y + Y\,m^2\,c^2)}{m^2\,c^2(X^{2} + Y^{2}) + (X\,P_y - Y\,P_x)^2}
\end{equation}
\begin{equation}
\{P_{x}, P_{y}\}_{DB}
= -\frac{\bigl(P_x^{2} + P_y^{2} + m^{2}c^{2}\bigr)\bigl(X\,P_y - \Omega\,Y\,P_x\bigr)}
       {m^{2}c^{2}\bigl(X^{2} + Y^{2}\bigr) + \bigl(X\,P_y - Y\,P_x\bigr)^{2}}
\end{equation}

By inspecting the brackets we observe that they all share the same denominator, and the brackets of the dynamical coordinates and the momenta also exhibit analogous structures in their numerators. These brackets coincide with those obtained in both the non-relativistic case and the reduced relativistic case; nevertheless, to arrive at that form, we must invoke the constraints enforced strongly to zero. In what follows, we only use the constraint \(\phi_{3}\),
\begin{equation}
  \phi_{3}\;=\;X P_{x}\;+\;Y P_{y}\;=\;0 ,
\end{equation}
which can be solved as
\begin{equation}
  \label{B2}
  X P_{x}\;=\;-\,Y P_{y}\; .
\end{equation}
Taking the square of \(\phi_{3}\) we find the following.
\begin{align}
  \label{B3}
  \phi_{3}^{\,2}
    &= P_{x}^{2}X^{2}+2P_{x}P_{y}XY+P_{y}^{2}Y^{2}=0 ,\\
  \label{B4}
 & \Rightarrow\quad
  P_{x}^{2}X^{2}+P_{y}^{2}Y^{2}
    = -\,2P_{x}P_{y}XY .
\end{align}
With these results, we proceed to examine the denominator, specifically the term \((X P_{y}-Y P_{x})^{2}\):
\begin{equation}
  (X P_{y}-Y P_{x})^{2}
    = P_{x}^{2}Y^{2}-2P_{x}P_{y}XY+P_{y}^{2}X^{2}.
    \label{B5}
\end{equation}
We now define
\begin{equation}
  (X^{2}+Y^{2})\bigl(P_{x}^{2}+P_{y}^{2}\bigr)
    = P_{x}^{2}X^{2}+P_{y}^{2}Y^{2}
      +P_{x}^{2}Y^{2}+P_{y}^{2}X^{2}.
     \label{B6}
\end{equation}
Substituting \eqref{B4} into \eqref{B6}, yields to
\begin{equation}
  (X^{2}+Y^{2})\bigl(P_{x}^{2}+P_{y}^{2}\bigr)
    = -\,2P_{x}P_{y}XY
      +P_{x}^{2}Y^{2}+P_{y}^{2}X^{2}.
        \label{B7}
\end{equation}
Comparing \eqref{B7} with \eqref{B5} immediately shows that
\begin{equation}
  (X P_{y}-Y P_{x})^{2}
    =(X^{2}+Y^{2})\bigl(P_{x}^{2}+P_{y}^{2}\bigr).
      \label{B8}
\end{equation}
The identity \eqref{B8} allows us to rewrite the common denominator of the brackets as
\begin{equation}
  \frac{1}{\bigl(X^{2}+Y^{2}\bigr)\bigl(P_{x}^{2}+P_{y}^{2}+m^{2}c^{2}\bigr)}.
    \label{B9}
\end{equation}
Using \eqref{B9}, we immediately obtain the Dirac bracket between the momenta.
\begin{equation}
  \{P_{x},P_{y}\}_{DB}
    = -\,\frac{X P_{y}-Y\Omega P_{x}}{X^{2}+Y^{2}}.
    \label{B10}
\end{equation}
For the Dirac bracket between the coordinates and the momenta, we must now analyze the numerator.
\begin{equation}
  Y^2 P_{x}^{2}-YX P_{x}P_{y}+Y^2 m^{2}c^{2}.
    \label{B11}
\end{equation}
Using \eqref{B2}, one readily sees that
\begin{equation}
  Y\bigl(Y P_{x}^{2}-X P_{x}P_{y}+Y m^{2}c^{2}\bigr)
    = Y^{2}\bigl(P_{x}^{2}+P_{y}^{2}+m^{2}c^{2}\bigr).
    \label{B12}
\end{equation}
Equation \eqref{B12} is the numerator of the bracket \(\{x,P_{x}\}_{DB}\); the same procedure applies to the numerators of the remaining brackets. Combining \eqref{B12} with
\eqref{B9}, we arrive at
\begin{equation}
  \{x,P_{x}\}_{DB}
    = \frac{Y^{2}}{X^{2}+Y^{2}}.
    \label{B13}
\end{equation}
Hence, we have shown that both the reduced and the extended versions of the constraints produce the same Dirac brackets. The key difference is that the extended case leads to considerably more cumbersome expressions containing redundant terms, which are removed once the constraints are imposed strongly.

Lastly, equations~(\ref{B14}--\ref{B18}) display the bracket values and the dynamical evolution associated with the variable~$\lambda$. These results are equivalent for both the Dirac constraint formalism and the Hamilton-Jacobi formalism.

\begin{equation}
\begin{aligned}
\{\lambda, P_x\} =\; & 
- \frac{c^2 P_y}{H_0^2 D^2} 
\left( X P_y - \Omega Y P_x \right) 
\left( H_0 + \frac{\theta}{2m} D \right) \\
& - \frac{\theta}{m D^2} X Y^2 (\Omega - 1)
\end{aligned}
\label{B14}
\end{equation}

\begin{equation}
\begin{aligned}
\{\lambda, P_y\} =\; & 
\frac{c^2 P_x}{H_0^2 D^2} 
\left( X P_y - \Omega Y P_x \right) 
\left( H_0 + \frac{\theta}{2m} D \right) \\
& + \frac{\theta}{m D^2} X^2 Y (\Omega - 1)
\end{aligned}
\label{B15}
\end{equation}

\begin{align}
\{\lambda, x\} = & 
- \frac{X c^2}{H_0^2 D^2} \Bigg[
\frac{\theta}{2m} D \,\Pi\notag \\ 
&+ H_0 (X P_y - \Omega Y P_x)
\Bigg]
\label{B16}
\end{align}
\begin{align}
\{\lambda, y\} = & 
\frac{Y c^2}{H_0^2 D^2} \Bigg[
\frac{\theta}{2m} D \,\Pi \notag \\
 &+ H_0 (X P_y - \Omega Y P_x)
\Bigg]
\label{B17}
\end{align}
\begin{align}
\dot{\lambda} =\; & 
\frac{c^2 (\Omega - 1)}{H_0^2 D^2} \Bigg[
\frac{\theta H_0}{m} X Y \,\Pi \notag \\
& + c^2 P_x P_y (X P_y - \Omega Y P_x)
\Bigg]
\label{B18}
\end{align}
\section{Appendix: Computer implementation in MATLAB for the calculations of results}
\label{AppendixC}

The Dirac constraint formalism—also known as the \emph{Dirac–Bergmann algorithm}—provides a systematic scheme for analysing singular systems by identifying and classifying their constraints.  
Because it is, in essence, an algorithm composed of well-defined steps, it is natural to transfer this procedure to a computational implementation that automates the treatment of such systems.

We employed \textsc{MATLAB}, an environment for numerical and symbolic computation primarily oriented toward matrix algebra. Since the Dirac formalism relies heavily on matrix operations, implementing the program in this environment is a natural choice, especially given the inclusion of the \emph{Symbolic Math Toolbox}.

This add-on enables symbolic computation and offers built-in functions for most of the algebraic operations required (e.g., functional derivatives, simplification, substitution, and determinant calculation).  
It also allows the creation of user-defined functions that encapsulate complex mathematical operations in several subprocesses, similar to packages in \textsc{Mathematica} \cite{romero2025singular}.

In our case, the native functions of the \emph{Symbolic Math Toolbox} cover all the fundamental operators involved in the Dirac formalism.  
Therefore, our strategy was to develop a library of functions that executes—sequentially and fully automatically—each stage of the Dirac–Bergmann algorithm.  
The user only needs to supply two inputs: the system’s Lagrangian and the set of canonical variables.  
From these, the code:

\begin{enumerate}
  \item Computes the Hessian matrix and its null vectors.
  \item Calculates the canonical momenta and detects the primary constraints.
  \item Builds the Hamiltonian and applies consistency conditions to obtain the remaining constraints.
  \item Classifies the constraints into first and second class.
  \item Generates the Poisson-bracket matrix of constraints and its inverse.
  \item Generates the canonical Dirac brackets.
  \item Produces the equations of motion for the canonical coordinates and momenta, \(\dot{q}_i\) and \(\dot{P}_i\).
\end{enumerate}

In this way, we obtain a flexible and reusable platform that facilitates the analysis of singular systems without the need to perform each step of the theoretical procedure manually, thereby reducing the risk of algebraic errors and accelerating the workflow.

\subsection{Function Structure in \textsc{MATLAB}}

In \textsc{MATLAB}, a typical function can, for didactic purposes, be broken down into three fundamental blocks:

\begin{enumerate}
  \item \textbf{Header and input interface}\\
        Defines the \emph{input arguments}, the \emph{output arguments}, and the function name.  
        The values the user wishes to process are supplied here, and it is possible to accept both required and optional inputs via mechanisms such as \verb|varargin| and \verb|Name,Value| pairs.  
        \begin{lstlisting}[language=Matlab,basicstyle=\ttfamily\small]
function [out1,out2] = myFunction(in1,in2,varargin)
        \end{lstlisting}

        These functions allow multiple inputs and outputs.

  \item \textbf{Body or processing section}\\
        Contains the sequence of instructions that transform the input data.  
        Here one can:
        \begin{itemize}
            \item Declare local and persistent variables.
            \item Invoke other functions or \textit{scripts}.
            \item Use control structures (\verb|if|, \verb|for|, \verb|switch|, etc.).
            \item Implement error handling with \verb|try/catch|.
        \end{itemize}
        The entire logical flow of the algorithm is described in this block.

  \item \textbf{Output assignment and closure}\\
        The final results are assigned to the output variables specified in the header.  
        These outputs are returned to the workspace that called the function; the user can store them, display them, or use them in subsequent computations.
        \begin{lstlisting}[language=Matlab,basicstyle=\ttfamily\small]
out1 = resultadoPrincipal;
out2 = diagnosticos;
end
        \end{lstlisting}
\end{enumerate}


%
%
%

This basic structure is used in the creation of all functions in the code base.

\subsection{SolveSingular function}

The centerpiece of the library is the \texttt{SolveSingular} function, which condenses the full implementation of Dirac’s constraint formalism for a user-provided Lagrangian.  
Its effectiveness lies in acting as an \emph{orchestrator}: it delegates each stage of the algorithm to sub-routines specifically designed for a single task (canonical-momentum calculation, detection of primary constraints, temporal-consistency checking, classification into first- and second-class constraints, construction of Dirac brackets, etc.).

\begin{lstlisting}[language=Matlab,basicstyle=\ttfamily\small]
[HE, R, W, DB] = SolveSingular(L, varargin)
\end{lstlisting}

To use it, the user only needs to supply:

\begin{itemize}
  \item \textbf{\(L\)} — the system’s Lagrangian, provided as a symbolic expression.
  \item \textbf{\texttt{varargin}} — the list of canonical coordinates, passed as comma-separated strings; e.g.\ \verb|'x', 'y', 'z'|.
\end{itemize}

The function automatically returns:

\begin{itemize}
  \item \(HE\): the extended Hamiltonian;
  \item \texttt{R}: the complete set of constraints found;
  \item \texttt{W}: the Poisson matrix of constraints;
  \item \texttt{DB}: the Dirac-bracket matrix for all canonical variables.
\end{itemize}

Moreover, the \texttt{SolveSingular} code incorporates several \verb|disp| calls that print to the console every step of the algorithm, the intermediate operations performed, and any potential issues encountered.  
This instrumentation was designed so that the user has real-time feedback on the calculation flow and can readily understand, debug, or document everything the program is executing, in Appendix \ref{AppendixD}, an example of the program’s output is shown. The code has been thoroughly validated across a wide variety of systems reported in the literature, consistently reproducing the expected results. Although the expressions we obtain are not identical to those published by each author (owing to differences in simplification criteria and computational techniques)our approach prioritizes a general and rigorous implementation, strictly following every step of the formalism without omissions.

As a result, the program accurately identifies singular systems of:

\begin{itemize}
  \item \textbf{First class}: containing only first-class constraints;
  \item \textbf{Second class}: containing only second-class constraints;
  \item \textbf{Mixed type}: involving both first- and second-class constraints.
\end{itemize}
In all cases, the algorithm automatically computes the equations of motion and provides a self-consistent set of results.
\newpage
\onecolumn

\newpage
Table.~\ref{tab:desc-lag} lists the benchmark systems used to verify the robustness and reproducibility of the program.

\begin{table}[htpb]
    \centering
    \renewcommand{\arraystretch}{1.4705} 
    %
    \resizebox{\textwidth}{!}{%
    \begin{tabular}{|p{3.2cm}|l|}
        \hline
        \textbf{Name} & \textbf{Lagrangian $L$} \\ \hline
        Brown System \cite{brown2022singular} &
        $\displaystyle 
        L(x,y,z,h) =\tfrac12\big[(x+\dot y+\dot z)^2 + (\dot h-\dot y)^2 + (x+2y)(x+2h)\big]$
        \\ \hline
        Compound spring \cite{brown2023singular,paulin2024singular} &
        $\displaystyle
        L(x,y) =\tfrac{m}{2}(\dot x + \dot y)^{2}
        + m g\,(x + y)
        - \tfrac{k_{1}}{2}(x - l_{1})^{2}
        - \tfrac{k_{2}}{2}(y - l_{2})^{2}$
        \\ \hline
        Pendulum and two springs \cite{brown2023singular,romero2025singular} &
        $\displaystyle
        L(x,y,\theta) =\tfrac{m}{2}\bigl(\dot x^{2} + \dot y^{2} + l^{2}\dot\theta^{2}\bigr)
        + ml(\dot x\cos\theta + \dot y\sin\theta)\dot\theta
        - mg\,(y - l\cos\theta) - k\,(x^{2} + y^{2} + d^{2})$
        \\ \hline
        Masses, springs and rings \cite{brown2023singular,romero2025singular} &
        $\displaystyle
        \begin{aligned}[t]
            L(\theta,\omega,\psi) =&\tfrac{mR^{2}}{2}\bigl(\dot\theta^{2}+\dot\omega^{2}+\dot\psi^{2}\bigr)
            - \tfrac{k}{2}
            \Bigl[(x-R\cos\theta)^{2}+(y-R\sin\theta)^{2}\\
            &+(x-R\cos\omega)^{2}+(y-R\sin\omega)^{2}
              +(x-R\cos\psi)^{2}+(y-R\sin\psi)^{2}\Bigr]
        \end{aligned}$
        \\ \hline
        Masses, rods, and springs \cite{brown2023singular} &
       $\displaystyle
\begin{aligned}[t]
    L(x,y,z,h) =\tfrac{m}{2} \Bigl[
        \left( \tfrac{\dot{x} + \dot{y}}{2} \right)^2
        + \left( \tfrac{\dot{y} + \dot{z}}{2} \right)^2
        + \left( \tfrac{\dot{z} + \dot{h}}{2} \right)^2
        + \left( \tfrac{\dot{h} + \dot{x}}{2} \right)^2
    \Bigr]
    - V(x, y, z, h)
\end{aligned}$
        \\ \hline
        Pairs of pulleys \cite{romero2025singular} &
        $\displaystyle
        L(x,y,z) =\tfrac{mR^{2}}{8}\bigl[(\dot x-\dot y)^2 + (\dot y-\dot z)^2 + (\dot z-\dot x)^2\bigr]
        - \tfrac{kR^{2}}{8}\bigl[(x-y)^2 + (y-z)^2 + (z-x)^2\bigr]$
        \\ \hline   
        Hyperboloid \cite{najafizade2022dirac} &
        $\displaystyle
        L(x,y,\lambda) =\tfrac{m}{2}\bigl(\dot x^{2}+\dot y^{2}-\dot z^{2}\bigr)
        - \lambda\Bigl(\tfrac{x^{2}+y^{2}}{\eta^{2}} - \tfrac{z^{2}}{1-\eta^{2}} - a^{2}\Bigr)$
        \\ \hline
        
        General Conic system \cite{barbosa2018gauge} &
        $\displaystyle
        L(x,y,z) =\tfrac12 m(\dot x^{2}+\dot y^{2})
        + z\Bigl(\tfrac12 A x^{2} + \tfrac12 B y^{2} + Cxy + Dx + Ey + F\Bigr)$
        \\ \hline
        
        Chern-Simons quantum mechanics \cite{bertin2014involutive} &
        $\displaystyle 
        L(x,y,q) =vq + \tfrac12 B\bigl[x(\dot y - qx) - y(\dot x + qy)\bigr]$
        \\ \hline

        The Christ-Lee model \cite{bertin2014involutive} &
        $\displaystyle 
        L(x, y, q) = 
        \tfrac{1}{2}(\dot{x}^2 + \dot{y}^2)
        - q(x \dot{y} - y \dot{x})
        + \tfrac{1}{2} q^2 (x^2 + y^2)
        - V(x^2 + y^2).$
        \\ \hline
        
        Particle motion on a torus \cite{gomez2024symplectic} &
        $\displaystyle
        \begin{aligned}[t]
            &L(\eta,\theta,\phi) =\tfrac{\mu}{2}\dot\eta^{2}
            + \tfrac{\mu}{2}\eta^{2}\dot\theta^{2}
            + \tfrac{\mu}{2}\bigl(b+\eta\sin\theta\bigr)^{2}\dot\phi^{2} 
             -\, V(\eta,\theta,\phi) - l\,(\eta - a)
        \end{aligned}$
        \\ \hline
        A partially integrable system \cite{Muslih2003} &
        $\displaystyle
        L(x,y) =\tfrac{1}{2}(\dot{x}^2 +\tfrac{1}{y^2})$
        \\ \hline
        A completely integrable system \cite{Muslih2003} &
        $\displaystyle
        L(x,y) =\tfrac{1}{2}\,m\,\omega\,(x\,\dot{y} - \dot{x}\,y - \omega\,(x^2 
         +y^2))$
        \\ \hline
        BĂleanu system \cite{bualeanu2001hamilton} &
        $\displaystyle
        L(x,y,z) =\tfrac{1}{2}\bigl(\dot{x}^{2} + \dot{y}^{2}\bigr)
        + \tfrac{1}{2}\,z^{2}\!\bigl(x^{2} + y^{2}\bigr)
        - z\,\bigl(x\,\dot{y} - y\,\dot{x}\bigr)
        - \tfrac{1}{2}\bigl(x^{2} + y^{2}\bigr).$
        \\ \hline
        The multi-dimensional rotator \cite{rothe2003hamilton, bertin2008non}&
        $\displaystyle
        L(x,y, \lambda) =\tfrac{1}{2}\bigl(\dot{x}^{2} + \dot{y}^{2}\bigr)
        + \lambda \,{\left(x\,\dot{x} +y\,\dot{y} \right)}.$
        \\ \hline
        Landau model \cite{rothe2003hamilton, bertin2008non} &
        $\displaystyle
        L(x,y) = \tfrac{1}{2}\left[B\,(x\,\dot{y} - \dot{x}\,y) -k\,(x^2 + y^2)\right]$
        \\ \hline
        The Güler’s example \cite{rabei1992hamilton, bertin2008non} &
        $\displaystyle
        L(x,y,z) =\tfrac{1}{2}\,\dot{x}^2 - \tfrac{1}{4}\,(\dot{y} - \dot{z})^2 + (x+z)\,\dot{y} - (x + y + z^2)$
        \\ \hline
        First-class Example \cite{nawafleh2004hamilton} &
        $\displaystyle
        L(x,y,z)=\tfrac{1}{2}\,(\dot{x}^2 + \dot{z}^2) +\dot{x}\,\dot{y} + y\,\dot{y} - x - z $
        \\ \hline
        First-class Example 2 \cite{nawafleh2004hamilton} &
        $\displaystyle
        L(x,y,z)=\tfrac{1}{2}(\dot{x}^2 + \dot{y}^2 + 2\dot{z}^2)
        + (\dot{x} + \dot{y} - x)\dot{z} + \tfrac{1}{2} x^2. $
        \\ \hline
        Mixed-class Example \cite{nawafleh2004hamilton} &
        $\displaystyle
        L(x,y,z,h)=\tfrac{1}{2}(\dot{x}^2 + \dot{y}^2 + \dot{z}^2 + \dot{h}^2)
        + \dot{x} \dot{h} + \dot{y} \dot{z} + x \dot{x} + \tfrac{1}{2} z^2. $
        \\ \hline
        Gauge-system Example \cite{golovnev2023role} &
        $\displaystyle
        L(x,y) = \tfrac{1}{2}(\dot{x}^2 -x^2\,(y-x)). $
        \\ \hline
        The spin one-half particle.  \cite{Deriglazov2017S} &
        $\displaystyle
        L(\omega,g,\phi) = \tfrac{1}{2g} \dot{\omega}^2 + g \tfrac{b^2}{2a^2} + \tfrac{1}{\phi}(\omega^2 - a^2). $
        \\ \hline
    \end{tabular}}
    \caption{Systems tested in the program}
    \label{tab:desc-lag}
\end{table}

\newpage
\section{Appendix: Example of output from the MATLAB code for the non-relativistic conical system}
\label{AppendixD}

The content of the gray boxes corresponds to the inputs in the MATLAB Live Script, while the remainder displays the output (Some outputs were removed because they were too extensive.) automatically generated by the code. In these blocks, it is sufficient to:
\begin{enumerate}
    \item Define the symbolic expressions (declared with \texttt{syms}) comprising the system’s Lagrangian \(L\).
    \item Apply the \texttt{solveSingular} function together with the corresponding canonical coordinates.
\end{enumerate}
In this way, the program implements the Dirac constraint formalism and automatically determines the full dynamics of the system.
\,

\begin{matlabcode}
clear

syms x y x_dot y_dot lambda m a_y Omega omega a_x e u real

L = (1/2)*m*(x_dot^2 + y_dot^2) 
+ lambda*((y - a_y)^2 + Omega*x^2 - 2*(omega)*x + a_x^2 - u^2*e^2)
\end{matlabcode}
\,
\begin{matlabsymbolicoutput}
L = 
\hskip1em $\displaystyle \lambda \,{\left({a_x }^2 +{a_y }^2 -2\,a_y \,y-e^2 \,u^2 +\Omega \,x^2 -2\,\omega \,x+y^2 \right)}+\frac{m\,{\left({\dot{x} }^2 +{\dot{y} }^2 \right)}}{2}$
\end{matlabsymbolicoutput}
\,
\begin{matlabcode}
[HH, R, WW, DB] = SolveSingular(L,'x', 'y', 'lambda');
\end{matlabcode}
\,
\begin{matlaboutput}
Hessian matrix:
\end{matlaboutput}
\,
\begin{matlabsymbolicoutput}
\hskip1em $\displaystyle \left(\begin{array}{ccc}
m & 0 & 0\\
0 & m & 0\\
0 & 0 & 0
\end{array}\right)$
\,
\end{matlabsymbolicoutput}
\,
\begin{matlaboutput}
Hessian determinant: 
\end{matlaboutput}
\,
\begin{matlabsymbolicoutput}
\hskip1em $\displaystyle 0$
\end{matlabsymbolicoutput}
\,
\begin{matlaboutput}
Hessian is singular.
--- Canonical momenta P_i ---
\end{matlaboutput}
\,
\begin{matlabsymbolicoutput}
\hskip1em $\displaystyle P_x = m\,\dot{x} $
\end{matlabsymbolicoutput}
\,
\begin{matlabsymbolicoutput}
\hskip1em $\displaystyle P_y = m\,\dot{y} $
\end{matlabsymbolicoutput}
\,
\begin{matlabsymbolicoutput}
\hskip1em $\displaystyle P_\lambda = 0$
\end{matlabsymbolicoutput}
\,
\begin{matlaboutput}
Null vector:
\end{matlaboutput}
\,
\begin{matlabsymbolicoutput}
\hskip1em $\displaystyle \left(\begin{array}{c}
0\\
0\\
1
\end{array}\right)$
\,
\end{matlabsymbolicoutput}
\,
\begin{matlaboutput}
Primary constraints:
\end{matlaboutput}
\,
\begin{matlabsymbolicoutput}
\hskip1em $\displaystyle P_{\lambda } $
\end{matlabsymbolicoutput}
\,
\begin{matlaboutput}
Canonical Hamiltonian obtained:
\end{matlaboutput}
\,
\begin{matlabsymbolicoutput}
\hskip1em $\displaystyle \frac{{P_x }^2 }{m}-\lambda \,{\left({a_x }^2 +{a_y }^2 -2\,a_y \,y-e^2 \,u^2 +\Omega \,x^2 -2\,\omega \,x+y^2 \right)}-\frac{{P_x }^2 -{P_y }^2 }{2\,m}$
\end{matlabsymbolicoutput}
\,
\begin{matlaboutput}
--- All constraints ---
Constraint 1:
\end{matlaboutput}
\,
\begin{matlabsymbolicoutput}
\hskip1em $\displaystyle P_{\lambda } $
\end{matlabsymbolicoutput}
\,
\begin{matlaboutput}
Constraint 2:
\end{matlaboutput}
\,
\begin{matlabsymbolicoutput}
\hskip1em $\displaystyle {a_x }^2 +{a_y }^2 -2\,a_y \,y-e^2 \,u^2 +\Omega \,x^2 -2\,\omega \,x+y^2 $
\end{matlabsymbolicoutput}
\,
\begin{matlaboutput}
Constraint 3:
\end{matlaboutput}
\,
\begin{matlabsymbolicoutput}
\hskip1em $\displaystyle -\frac{2\,{\left(P_y \,a_y +P_x \,\omega -P_y \,y-\Omega \,P_x \,x\right)}}{m}$
\end{matlabsymbolicoutput}
\,
\begin{matlaboutput}
Constraint 4:
\end{matlaboutput}
\,
\begin{matlabsymbolicoutput}
\hskip1em $\displaystyle \frac{2\,{\left(2\,\lambda \,m\,\Omega^2 \,x^2 +\Omega \,{P_x }^2 -4\,\lambda \,m\,\Omega \,\omega \,x+{P_y }^2 +2\,\lambda \,m\,{a_y }^2 -4\,\lambda \,m\,a_y \,y+2\,\lambda \,m\,\omega^2 +2\,\lambda \,m\,y^2 \right)}}{m^2 }$
\end{matlabsymbolicoutput}
\,

\begin{matlaboutput}
Initial W matrix:
\end{matlaboutput}
\,
\begin{matlabsymbolicoutput}
\hskip1em $\displaystyle \begin{array}{l}
\left(\begin{array}{cccc}
0 & 0 & 0 & -\sigma_1 \\
0 & 0 & \sigma_1  & -\frac{8\,{\left(-P_x \,x\,\Omega^2 +P_x \,\omega \,\Omega +P_y \,a_y -P_y \,y\right)}}{m^2 }\\
0 & -\sigma_1  & 0 & \sigma_2 \\
\sigma_1  & \frac{8\,{\left(-P_x \,x\,\Omega^2 +P_x \,\omega \,\Omega +P_y \,a_y -P_y \,y\right)}}{m^2 } & -\sigma_2  & 0
\end{array}\right)\\
\,
\mathrm{}\\
\textrm{where}\\
\mathrm{}\\
\;\;\sigma_1 =\frac{4\,{\left(\Omega^2 \,x^2 -2\,\Omega \,\omega \,x+{a_y }^2 -2\,a_y \,y+\omega^2 +y^2 \right)}}{m}\\
\mathrm{}\\
\;\;\sigma_2 =\frac{8\,{\left(-2\,\lambda \,m\,\Omega^3 \,x^2 +\Omega^2 \,{P_x }^2 +4\,\lambda \,m\,\Omega^2 \,\omega \,x-2\,\lambda \,m\,\Omega \,\omega^2 +{P_y }^2 -2\,\lambda \,m\,{a_y }^2 +4\,\lambda \,m\,a_y \,y-2\,\lambda \,m\,y^2 \right)}}{m^3 }
\end{array}$
\,
\end{matlabsymbolicoutput}
\,
\begin{matlaboutput}
Null vectors of W:
     0

Null vectors empty/trivial → original constraints are kept.
Second-class constraints:
\end{matlaboutput}
\,
\begin{matlabsymbolicoutput}
\hskip1em $\displaystyle \left(\begin{array}{c}
P_{\lambda } \\
{a_x }^2 +{a_y }^2 -2\,a_y \,y-e^2 \,u^2 +\Omega \,x^2 -2\,\omega \,x+y^2 \\
-\frac{2\,{\left(P_y \,a_y +P_x \,\omega -P_y \,y-\Omega \,P_x \,x\right)}}{m}\\
\frac{2\,{\left(2\,\lambda \,m\,\Omega^2 \,x^2 +\Omega \,{P_x }^2 -4\,\lambda \,m\,\Omega \,\omega \,x+{P_y }^2 +2\,\lambda \,m\,{a_y }^2 -4\,\lambda \,m\,a_y \,y+2\,\lambda \,m\,\omega^2 +2\,\lambda \,m\,y^2 \right)}}{m^2 }
\end{array}\right)$
\,
\end{matlabsymbolicoutput}
\,
\begin{matlaboutput}
Degrees of freedom: 1
Inverse of the W matrix:
\end{matlaboutput}
\begin{matlabsymbolicoutput}
\hskip1em $\displaystyle \begin{array}{l}
\left(\begin{array}{cccc}
0 & \sigma_1  & \sigma_2  & \frac{m}{4\,\sigma_3 }\\
-\sigma_1  & 0 & -\frac{m}{4\,\sigma_3 } & 0\\
-\sigma_2  & \frac{m}{4\,\sigma_3 } & 0 & 0\\
-\frac{m}{4\,\sigma_3 } & 0 & 0 & 0
\end{array}\right)\\
\,
\mathrm{}\\
\textrm{where}\\
\mathrm{}\\
\;\;\sigma_1 =\frac{-2\,\lambda \,m\,\Omega^3 \,x^2 +\Omega^2 \,{P_x }^2 +4\,\lambda \,m\,\Omega^2 \,\omega \,x-2\,\lambda \,m\,\Omega \,\omega^2 +{P_y }^2 -2\,\lambda \,m\,{a_y }^2 +4\,\lambda \,m\,a_y \,y-2\,\lambda \,m\,y^2 }{2\,m\,{\sigma_3 }^2 }\\
\mathrm{}\\
\;\;\sigma_2 =\frac{-P_x \,x\,\Omega^2 +P_x \,\omega \,\Omega +P_y \,a_y -P_y \,y}{2\,{\sigma_3 }^2 }\\
\mathrm{}\\
\;\;\sigma_3 =\Omega^2 \,x^2 -2\,\Omega \,\omega \,x+{a_y }^2 -2\,a_y \,y+\omega^2 +y^2 
\end{array}$
\,
\end{matlabsymbolicoutput}
\,
\begin{matlaboutput}
---------------------------------------------------------------
Equations of motion (Poisson, extended Hamiltonian):
\dot{x} = 
\end{matlaboutput}
\,
\begin{matlabsymbolicoutput}
\hskip1em $\displaystyle \frac{P_x \,m+4\,\Omega \,P_x \,u_4 -2\,m\,\omega \,u_3 +2\,\Omega \,m\,u_3 \,x}{m^2 }$
\end{matlabsymbolicoutput}
\,
\begin{matlaboutput}
 
\dot{y} = 
\end{matlaboutput}
\,
\begin{matlabsymbolicoutput}
\hskip1em $\displaystyle \frac{P_y \,m+4\,P_y \,u_4 -2\,a_y \,m\,u_3 +2\,m\,u_3 \,y}{m^2 }$
\end{matlabsymbolicoutput}
\,
\begin{matlaboutput}
 
\dot{lambda} = 
\end{matlaboutput}
\,
\begin{matlabsymbolicoutput}
\hskip1em $\displaystyle u_1 $
\end{matlabsymbolicoutput}
\,
\begin{matlaboutput}
 
\dot{P_x} = 
\end{matlaboutput}
\,
\begin{matlabsymbolicoutput}
\hskip1em $\displaystyle -\frac{2\,{\left(\Omega \,P_x \,u_3 +\lambda \,m\,\omega -4\,\Omega \,\lambda \,\omega \,u_4 -\Omega \,\lambda \,m\,x+4\,\Omega^2 \,\lambda \,u_4 \,x\right)}}{m}$
\end{matlabsymbolicoutput}
\,
\begin{matlaboutput}
 
\dot{P_y} = 
\end{matlaboutput}
\,
\begin{matlabsymbolicoutput}
\hskip1em $\displaystyle -\frac{2\,{\left(P_y \,u_3 +a_y \,\lambda \,m-4\,a_y \,\lambda \,u_4 -\lambda \,m\,y+4\,\lambda \,u_4 \,y\right)}}{m}$
\end{matlabsymbolicoutput}
\,
\begin{matlaboutput}
 
\dot{P_lambda} = 
\end{matlaboutput}
\,
\begin{matlabsymbolicoutput}
\hskip1em $\displaystyle -\frac{4\,u_4 \,{\left(\Omega^2 \,x^2 -2\,\Omega \,\omega \,x+{a_y }^2 -2\,a_y \,y+\omega^2 +y^2 \right)}}{m}$
\end{matlabsymbolicoutput}
\,
\begin{matlaboutput}
 
All constraints are SECOND class.
---------------------------------------------------------------
Showing canonical Dirac brackets
{x, lambda}_D = 
\end{matlaboutput}
\,
\begin{matlabsymbolicoutput}
\hskip1em $\displaystyle \frac{{\left(a_y -y\right)}\,{\left(P_y \,\omega -\Omega \,P_x \,a_y -\Omega \,P_y \,x+\Omega \,P_x \,y\right)}}{m\,{{\left(\Omega^2 \,x^2 -2\,\Omega \,\omega \,x+{a_y }^2 -2\,a_y \,y+\omega^2 +y^2 \right)}}^2 }$
\end{matlabsymbolicoutput}
\,
\begin{matlaboutput}
{x, P_x}_D = 
\end{matlaboutput}
\,
\begin{matlabsymbolicoutput}
\hskip1em $\displaystyle \frac{{{\left(a_y -y\right)}}^2 }{\Omega^2 \,x^2 -2\,\Omega \,\omega \,x+{a_y }^2 -2\,a_y \,y+\omega^2 +y^2 }$
\end{matlabsymbolicoutput}
\,
\begin{matlaboutput}
{x, P_y}_D = 
\end{matlaboutput}
\,
\begin{matlabsymbolicoutput}
\hskip1em $\displaystyle -\frac{{\left(\omega -\Omega \,x\right)}\,{\left(2\,a_y -2\,y\right)}}{2\,{\left(\Omega^2 \,x^2 -2\,\Omega \,\omega \,x+{a_y }^2 -2\,a_y \,y+\omega^2 +y^2 \right)}}$
\end{matlabsymbolicoutput}
\,
\begin{matlaboutput}
{y, lambda}_D = 
\end{matlaboutput}
\,
\begin{matlabsymbolicoutput}
\hskip1em $\displaystyle -\frac{{\left(\omega -\Omega \,x\right)}\,{\left(P_y \,\omega -\Omega \,P_x \,a_y -\Omega \,P_y \,x+\Omega \,P_x \,y\right)}}{m\,{{\left(\Omega^2 \,x^2 -2\,\Omega \,\omega \,x+{a_y }^2 -2\,a_y \,y+\omega^2 +y^2 \right)}}^2 }$
\end{matlabsymbolicoutput}
\,
\begin{matlaboutput}
{y, P_x}_D = 
\end{matlaboutput}
\,
\begin{matlabsymbolicoutput}
\hskip1em $\displaystyle -\frac{{\left(2\,\omega -2\,\Omega \,x\right)}\,{\left(a_y -y\right)}}{2\,{\left(\Omega^2 \,x^2 -2\,\Omega \,\omega \,x+{a_y }^2 -2\,a_y \,y+\omega^2 +y^2 \right)}}$
\end{matlabsymbolicoutput}
\,
\begin{matlaboutput}
{y, P_y}_D = 
\end{matlaboutput}
\,
\begin{matlabsymbolicoutput}
\hskip1em $\displaystyle \frac{{{\left(\omega -\Omega \,x\right)}}^2 }{\Omega^2 \,x^2 -2\,\Omega \,\omega \,x+{a_y }^2 -2\,a_y \,y+\omega^2 +y^2 }$
\end{matlabsymbolicoutput}
\,
\begin{matlaboutput}
{P_x, P_y}_D = 
\end{matlaboutput}
\,
\begin{matlabsymbolicoutput}
\hskip1em $\displaystyle \frac{P_y \,\omega -\Omega \,P_x \,a_y -\Omega \,P_y \,x+\Omega \,P_x \,y}{\Omega^2 \,x^2 -2\,\Omega \,\omega \,x+{a_y }^2 -2\,a_y \,y+\omega^2 +y^2 }$
\end{matlabsymbolicoutput}
\,
\begin{matlaboutput}
Zero Dirac brackets:
{x, y}, {x, P_lambda}, {y, P_lambda}, {lambda, P_lambda}, {P_x, P_lambda}, 
{P_y, P_lambda}
---------------------------------------------------------------
Reduced Hamiltonian:
\end{matlaboutput}
\,
\begin{matlabsymbolicoutput}
\hskip1em $\displaystyle \frac{{P_x }^2 }{m}-\frac{{P_x }^2 -{P_y }^2 }{2\,m}$
\end{matlabsymbolicoutput}
\,
\begin{matlaboutput}
---------------------------------------------------------------
Equations of motion (Dirac, all constraints):
\dot{x} = 
\end{matlaboutput}
\,
\begin{matlabsymbolicoutput}
\hskip1em $\displaystyle \frac{{\left(a_y -y\right)}\,{\left(P_x \,a_y -P_y \,\omega -P_x \,y+\Omega \,P_y \,x\right)}}{m\,{\left(\Omega^2 \,x^2 -2\,\Omega \,\omega \,x+{a_y }^2 -2\,a_y \,y+\omega^2 +y^2 \right)}}$
\end{matlabsymbolicoutput}
\,
\begin{matlaboutput}
 
\dot{y} = 
\end{matlaboutput}
\,
\begin{matlabsymbolicoutput}
\hskip1em $\displaystyle -\frac{{\left(\omega -\Omega \,x\right)}\,{\left(P_x \,a_y -P_y \,\omega -P_x \,y+\Omega \,P_y \,x\right)}}{m\,{\left(\Omega^2 \,x^2 -2\,\Omega \,\omega \,x+{a_y }^2 -2\,a_y \,y+\omega^2 +y^2 \right)}}$
\end{matlabsymbolicoutput}
\,
\begin{matlaboutput}
 
\dot{P_x} = 
\end{matlaboutput}
\,
\begin{matlabsymbolicoutput}
\hskip1em $\displaystyle \frac{P_y \,{\left(P_y \,\omega -\Omega \,P_x \,a_y -\Omega \,P_y \,x+\Omega \,P_x \,y\right)}}{m\,{\left(\Omega^2 \,x^2 -2\,\Omega \,\omega \,x+{a_y }^2 -2\,a_y \,y+\omega^2 +y^2 \right)}}$
\end{matlabsymbolicoutput}
\,
\begin{matlaboutput}
 
\dot{P_y} = 
\end{matlaboutput}
\,
\begin{matlabsymbolicoutput}
\hskip1em $\displaystyle -\frac{P_x \,{\left(P_y \,\omega -\Omega \,P_x \,a_y -\Omega \,P_y \,x+\Omega \,P_x \,y\right)}}{m\,{\left(\Omega^2 \,x^2 -2\,\Omega \,\omega \,x+{a_y }^2 -2\,a_y \,y+\omega^2 +y^2 \right)}}$
\end{matlabsymbolicoutput}
\,
\begin{matlaboutput}
 
\dot{P_lambda} = 
\end{matlaboutput}
\,
\begin{matlabsymbolicoutput}
\hskip1em $\displaystyle 0$
\end{matlabsymbolicoutput}

\end{document}